\documentclass[fleqn,usenatbib]{mnras}

\usepackage{newtxtext,newtxmath}

\usepackage[T1]{fontenc}

\DeclareRobustCommand{\VAN}[3]{#2}
\let\VANthebibliography\thebibliography
\def\thebibliography{\DeclareRobustCommand{\VAN}[3]{##3}\VANthebibliography}

\usepackage{fontawesome}

\usepackage{graphicx}	
\usepackage{amsmath}	

\title[Deconvolving Conditional Distributions]{Extreme Deconvolution Reimagined: Conditional Densities via Neural Networks and an Application in Quasar Classification}

\author[Y. K et al.]{
Yi Kang,$^{1}$\thanks{E-mail: yi\_kang@ucsb.edu}, Joseph F. Hennawi$^{1,2}$, Jan-Torge Schindler$^{3}$, John Tamanas$^{4}$, Riccardo Nanni$^{2}$, 
\\
$^{1}$Department of Physics, Broida Hall, University of California, Santa Barbara, CA 93106, USA\\
$^{2}$Leiden Observatory, Niels Bohrweg 2, Leiden, 2333 CA, the Netherlands\\
$^{3}$Hamburg Observatory, Gojenbergsweg 112, 21029 Hamburg, Germany\\
$^{4}$The Department of Astronomy and Astrophysics, University of California Santa Cruz, 1156 High Street, Santa Cruz, CA 95064, USA
}

\date{Accepted XXX. Received YYY; in original form ZZZ}

\pubyear{2024}

\begin{document}
\label{firstpage}
\pagerange{\pageref{firstpage}--\pageref{lastpage}}
\maketitle

\begin{abstract}
    Density estimation is a fundamental problem that arises in many areas of astronomy, with applications ranging from selecting quasars using color distributions to characterizing stellar abundances. Astronomical observations are inevitably noisy; however, the density of a noise-free feature is often the desired outcome. The extreme-deconvolution (XD) method can be used to deconvolve the noise and obtain noise-free density estimates by fitting a mixture of Gaussians to data where each sample has non-identical (heteroscedastic) Gaussian noise. However, XD does not generalize to cases where some feature dimensions have highly non-Gaussian distribution, and no established method exists to overcome this limitation. We introduce a possible solution using neural networks to perform Gaussian mixture modeling of the Gaussian-like dimensions conditioned on those non-Gaussian features. The result is the CondXD algorithm, a generalization of XD that performs noise-free conditional density estimation. We apply CondXD to a toy model and find that it is more accurate than other approaches. We further test our method on a real-world high redshift quasar versus contaminant classification problem. Specifically, we estimate noise-free densities in flux-ratio (i.e., color) space for contaminants, conditioned on their magnitude. Our results are comparable to the existing method, which divides the samples into magnitude bins and applies XD separately in each bin, and our method is approximately ten times faster. Overall, our method has the potential to significantly improve estimating conditional densities and enable new discoveries in astronomy.
\end{abstract}




\section{Introduction}

Density distribution estimation is an active area of research in astronomy, with a key attention on uncovering the underlying distributions of various astronomical properties. For example, \cite{Buder2022} used deconvolution techniques to estimate the distribution of the abundances of accreted stars, while \cite{Bovy2011, Mortlock2012, Bovy2012, Nanni2022} et al. applied similar methods to measure the flux distribution of quasars. Other researchers, such as \cite{Bird2021} and \cite{Ivezic2021}, have used density distribution estimation to infer the galactic structure and predict size estimations observed by the Rubin Observatory \citep{Ivezic2019}, respectively.

In practice, the physical attributes of astronomical targets are rarely measured without substantial and heteroscedastic uncertainties, and the estimation of the underlying distribution is never an easy task. The observations can be regarded as samples drawn from an (noiseless) underlying distribution convolved with the distribution of noise, thus the estimation of the underlying distribution is also referred to as deconvolution. While density deconvolution of noisy distributions has been extensively studied in the literature, early works such as \cite{Devroye1989, Stefanski1990, Zhang1990, Fan1991a, Fan1991b} often assumed that the distributions are univariate and the noise distribution is identical for every measurement, neglecting the heteroscedasticity of the measurements. Moreover, most of the early studies applied nonparametric approaches that cannot be implemented when samples with missing measurements (missing data) are encountered. 

To address these complications, \cite{Bovy2011XD} developed an extreme deconvolution (XD) algorithm that works for noisy, heterogeneous, and missing data\footnote{\faGithub\ \url{https://github.com/jobovy/extreme-deconvolution}}. They used a Gaussian Mixture Model (GMM) to fit the underlying distribution of a set of noisy samples. With the assumption of Gaussian underlying distribution and Gaussian noise distribution with zero mean, the noisy density distribution (i.e. convolution of the underlying and noise distribution) is equivalent to adding the covariance matrix of the noise distribution to that of the underlying distribution. Iteratively applying the expectation and maximization process to increase the likelihood of the noisy samples on the noisy distribution, the underlying distribution is estimated. As demonstrated in \cite{Bovy2011XD}, the performance of XD is capable of inferring the 3-D velocity distribution of stars around the Sun given the noisy 2-D, transverse velocity measurements from the \textit{Hipparcos} satellite. It is also robust in obtaining the optimal fit even given poor initialization. Recently, the scalable XD algorithm developed by \cite{sXD} improved the XD code with modern machine learning algorithms (e.g., stochastic gradient descent and mini-batches) to seek for the GMM best-fit parameters, instead of using an iterative expectation-maximization approach on the full data set. Similar studies were also conducted by \cite{Hosseini2015,Hosseini2017,Gepperth2019}.

Once pulished, XD has gained wide applications especially in the field of classifying quasars and contaminants. \cite{Bovy2011,White2012} used XD to model the distribution of quasars and stars from SDSS \citep{York_SDSS} in the (relative) flux space. With these distributions they evaluated the probabilities of sources to be a quasar or star based on their noisy measurements. Later, \cite{Bovy2012, Myers2015, DiPompeo2015} deployed this approach to model the distribution of optical, ultraviolet, and infrared band fluxes as well as redshifts of quasars, which yielded more accurate classification of quasars and capability of redshift estimation. In the era of high redshift observations facilitated by state-of-the-art telescopes like JWST and Euclid, \cite{Nanni2022} applied the XD method to distinguish high-redshift ($z>6$) quasars from contaminants. In their algorithm, one of the most crucial parts is to model the underlying distributions of quasars and contaminants fluxes, by deconvolving their noisy measurements. In both cases, the probability density of quasars will have a dominant power-law shape corresponding to the number counts as a function of apparent magnitude, which is hard to be approximated by Gaussian distributions. Therefore, they divided the samples into magnitude bins of the detection band to limit the variation within each bin, and applied the XD algorithm individually in every bin.

In fact, not all physical distributions are fixed, and some of them are dependent on certain variables, like the dependency of fluxes, colors and magnitudes on the luminosity or redshift of quasars, and the dependency of the 3-D velocity of stars on their Galactocentric distance and metallicity in the Milky Way. Both \cite{Bovy2011XD} and \cite{sXD} did not take into account of the dependencies of their models on other physical variables, while \cite{Bovy2011,Nanni2022,Bird2021} used the binning approach mentioned above. Alternatively, in theory, XD could still consider the variables as extra features (dimensions), estimate the general distribution, and then condition on these variables like in \cite{Bovy2012}. However, once the conditionals have significantly non-Gaussian marginal distributions, the general distribution would require a large number of Gaussians to be described.

Complicated density distributions can be modeled with a set of weighted basic density distributions, called density mixture. When the complicated distribution is also conditional (dependent on some variable), one can employ mixture density networks (\S 5.6 in \citealt{Bishop}) to fit the distribution. A mixture density network allows the parameters (e.g. for GMM, mixing coefficients\footnote{It is more often referred as `weights', but we use the term `mixing coefficients' to avoid confusion with the `weights' of the NN.}, means and covariance matrices) of the density mixture to be generated by a neural network (NN) that takes the conditionals as the input. This motivates us to combine the XD and mixture density networks in order to deconvolve the noisy conditional distributions of astronomical sources. In this paper, following \cite{sXD}, we use modern machine learning methods to find the optimal fit for noisy distributions using both a simple toy model and an astronomical real case classification scenario, demonstrating the capabilities of our conditional XD algorithm, CondXD, in deriving the underlying (noiseless) distribution from a noisy one. In \S \ref{sec:method} we provide a general description of the CondXD method. In \S \ref{sec:exp} we conduct an experiment to test the performance of CondXD on a simple toy model. Using the same toy model, in \S \ref{sec:bin} we compare the deconvolving capability of both the CondXD and a binning approach similar to the one from \cite{Nanni2022}. In \S \ref{sec:app} we apply CondXD to a realistic astronomy case and compare it with the binning method from \cite{Nanni2022}. In \S \ref{sec:conclution} we provide our conclusions and discussions.

\section{Method}
\label{sec:method}
In its standard form, XD estimates the underlying probability distribution $p(\mathbf{X})$ of a noisy sample using a GMM $\hat{p}(\mathbf{X} \mid \hat{\boldsymbol{\alpha}}, \hat{\boldsymbol{\mu}}, \hat{\mathbf{V}})$, defined by a set of $K$ mixing coefficients $\hat{\boldsymbol{\alpha}}$, means $\hat{\boldsymbol{\mu}}$ and covariance matrices $\hat{\mathbf{V}}$. We use hat notation to indicate estimated statistics, and leave out the notation to represent the true values of these quantities when they are generated from an underlying Gaussian mixture. However, if the probability density, $p$, is conditioned on a variable $\mathbf{c}$, such that $p(\mathbf{X}| \mathbf{c})$, the correct form for the estimator is: $\hat{p}(\mathbf{X}\mid \hat{\boldsymbol{\alpha}}(\mathbf{c}), \hat{\boldsymbol{\mu}}(\mathbf{c})), \hat{\mathbf{V}}(\mathbf{c}))$.  

In this work, we build our conditional GMM estimator using a NN with weights $\boldsymbol{\phi}$. In this case the GMM parameters become functions of $\boldsymbol{\phi}$ and conditional $\mathbf{c}$, written as: $\hat{\boldsymbol{\alpha}} (\boldsymbol{\phi}, \mathbf{c})$, $\hat{\boldsymbol{\mu}} (\boldsymbol{\phi},  \mathbf{c})$, $\hat{\mathbf{V}} (\boldsymbol{\phi}, \mathbf{c})$. Consequently, for notational simplicity we henceforth express the estimator as $\hat p(\mathbf{X}\mid \boldsymbol{\phi}, \mathbf{c})$. 

In \S \ref{subsec:arch} we describe the architecture of our NN, while in \S \ref{subsec:loss} we describe how we define the loss function of our method, and \S \ref{subsec:strategy} introduces the technical details implemented to improve training. Hereafter, we call our technique CondXD, whose code is available on Github\footnote{\faGithub~\url{https://github.com/enigma-igm/CondXD}}.

\begin{figure}
    \includegraphics[width=\columnwidth]{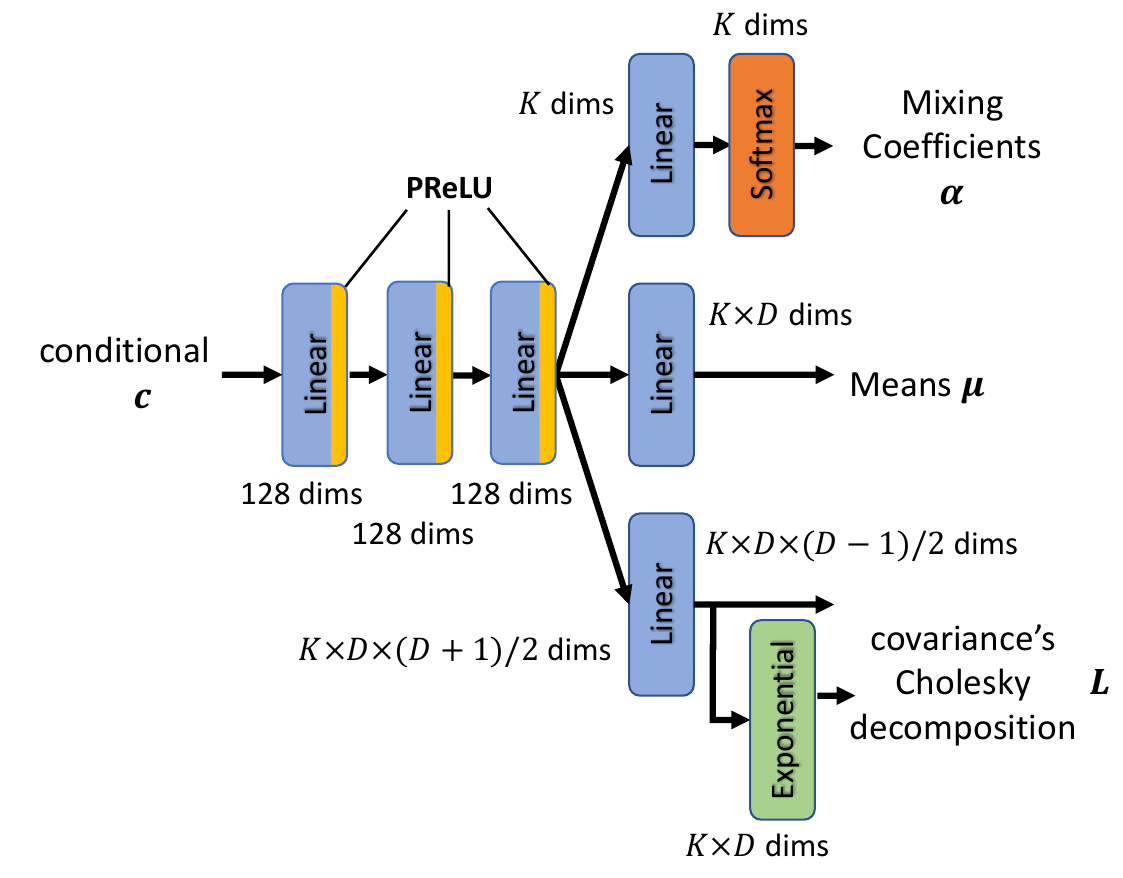}
    \caption{Schematic of the CondXD neural network. It takes in conditional $\mathbf{c}$ and outputs the parameters of a GMM, i.e. the mixing coefficients, means and Cholesky factors of the covariance matrices. Blocks are layers in the architecture, the types of which are indicated by annotations. The yellow ends refer to the \texttt{PReLU} activation functions after the current layer. $K$ is the number of Gaussians in the model, and $D$ is the dimension of the Gaussians, while both are hyperparameters. In practice, $D$ is automatically determined by the dimension of the samples. The dimensions of the outputs computed by every layer are also labeled.}
    \label{fig:model}
\end{figure}

\subsection{Architecture of the Neural Network}
\label{subsec:arch}
Our NN has a stem-branch structure shown in Figure \ref{fig:model}.  The stem constitutes a sequence of 
three linear layers, which branches off into three output layers for the mixing coefficients, means, and covariances, respectively. It takes in the conditional $\mathbf{c}$ and outputs the parameters of the GMM. In general, the density distribution of a noisy sample can depend on several variables, so that the conditional $\mathbf{c}$ is actually multi-dimensional. The number of Gaussians of the GMM ($K$) is a hyperparameter, and the dimension of the GMM ($D$) is determined by the dimension of the data from the observations, whose density is to be estimated. 

In the structure of the NN we simply use \texttt{Linear} layers everywhere. The \texttt{Linear} layers multiply the input with matrices, the elements of which are the weights $\boldsymbol{\phi}$ of our neural network. In the stem part, all the \texttt{Linear} layers are followed by the \texttt{PReLU} (Parametric Rectified Linear Unit; \citealt{PReLU}) activation functions. At the output layer for the mixing coefficients branch (see orange block in Figure \ref{fig:model}) we follow
standard practice, using a softmax activation to ensure that all mixing coefficients are positive and sum up to unity:
\begin{equation}
\hat\alpha_j = \frac{\exp(\beta_j)}{\sum_j\exp(\beta_j)}, 
\end{equation}
where $\hat\alpha_j$ is the mixing coefficient for the $j$th Gaussian, and $\beta_j$ is the $j$th output of the last \texttt{Linear} layer of the mixing coefficients branch. For the means branch we do not implement any further processing than a \texttt{Linear} layer, since there is not rigorous requirements on it. At last, instead of directly generating the covariances that have to satisfy symmetry and positive semi-definiteness, we again follow standard practice, generate the Cholesky decomposition factors $\hat{\mathbf{L}}(\boldsymbol{\phi}, \mathbf{c})$. The Cholesky decomposition is defined as:
\begin{equation}
    \hat{\mathbf{V}}_j = \hat{\mathbf{L}}_j \hat{\mathbf{L}}^\mathrm{T}_j,
    \label{eqn:covar}
\end{equation}
where $\hat{\mathbf{L}}_j$ is the $j$th Cholesky factor, which is a lower triangular matrix with shape $(D, D)$, and $\hat{\mathbf{V}}_j$ is the $j$th covariance matrix of the $K$ Gaussians. Each of the $K$ Cholesky factors has all zero values in the upper-right triangle, and all diagonal elements are positive for every matrix. In this way $\hat{\mathbf{V}}_j$ is guaranteed to be symmetric and positive semi-definite. In the covariance branch (see the green block in Figure~\ref{fig:model}), the diagonal elements of every Cholesky factor have been processed by an exponential activation function to ensure positivity.

With this architecture we generate all the GMM parameters. In practice, these parameters define the model that describes the deconvolved density distribution of the noisy samples.

\subsection{Loss Function}
\label{subsec:loss}
To train the NN, or in other words, to find the best-fit weights $\boldsymbol{\phi}$ in the NN, we need to quantify how well the GMM represents the data samples by utilizing a loss function. In our case, the Kullback–Leibler divergence (KL divergence or $D_\mathrm{KL}$; \citealt{D_KL}) is chosen as a standard practice to measure how different the GMM is from the underlying distribution. The KL divergence is defined as:
\begin{equation}
\begin{aligned}
     D_{\mathrm{KL}}\left(p \| \hat p\right)=\int  p(\mathbf{x} & \mid \mathbf{c};\textbf{S}(\mathbf{x})) \ln \left(\frac{p(\mathbf{x} \mid \mathbf{c};\textbf{S}(\mathbf{x}))}{\hat p(\mathbf{x} \mid \boldsymbol{\phi}; \mathbf{c};\textbf{S}(\mathbf{x}))}\right) \mathrm{d} \mathbf{x} \\
     =\int p(\mathbf{x} & \mid \mathbf{c}; \textbf{S}(\mathbf{x})) \ln p(\mathbf{x} \mid \mathbf{c};\textbf{S}(\mathbf{x})) \mathrm{~d} \mathbf{x} \\
     & - \int p(\mathbf{x} \mid  \mathbf{c};\textbf{S}(\mathbf{x})) \ln \hat p(\mathbf{x} \mid \boldsymbol{\phi}; \mathbf{c};\textbf{S}(\mathbf{x})) \mathrm{~d} \mathbf{x},
\end{aligned}
\label{eqn:KL}
\end{equation}
where $\mathbf{S}(\mathbf{x})$ is the noise covariance of the random sample $\mathbf{x}$. With the heteroscedastic assumption, each $\mathbf{x}$ can have its own noise, thus $\mathbf{S}$ can also be regarded as a function of $\mathbf{x}$, i.e. $\mathbf{S}(\mathbf{x})$. In practice, observed data are always noisy, thus both distributions $p$ and $\hat p$ in eqn. (\ref{eqn:KL}) have been convolved with noise. In the above equation, $p(\mathbf{x} \mid \mathbf{c};\textbf{S}(\mathbf{x}))$ is short for $p(\mathbf{X}=\mathbf{x} \mid \mathbf{c};\textbf{S}(\mathbf{x}))$, and $p(\mathbf{X} \mid \mathbf{c};\textbf{S}(\mathbf{x}))$ is the noise convolved underlying density $p(\mathbf{X} \mid \mathbf{c})$. Similarly, $\hat p(\mathbf{x}\mid \boldsymbol{\phi};\mathbf{c};\textbf{S}(\mathbf{x}))$ is the probability of $\mathbf{x}$ under the noise convolved GMM estimator for $p(\mathbf{X}\mid \mathbf{c};\textbf{S}(\mathbf{x}))$. As the samples $\mathbf{x}$ are from the noise convolved distribution $p(\mathbf{X} \mid \mathbf{c};\textbf{S}(\mathbf{x}))$, the integration in eqn. (\ref{eqn:KL}) is averaging over the sample space.

The goal is to find the model $\hat p (\mathbf{X}|\boldsymbol{\phi};\mathbf{c};\textbf{S}(\mathbf{x}))$ that minimizes the KL divergence with training set $\{\mathbf{x}, \mathbf{c}, \textbf{S}\}$. The first term in the second line of eqn. (\ref{eqn:KL}) is a constant that does not depend on the NN weights $\boldsymbol{\phi}$, hence only the second term needs to be minimized. Therefore, we can define the loss function as the second term, i.e. the negative of the log-probability of the model averaged over the underlying distribution. Since we do not have access to the noise convolved underlying distribution $p(\mathbf{X} \mid \mathbf{c}; \textbf{S})$ (this is what we are trying to estimate), but we do have access to noisy samples $\{\mathbf{x}, \mathbf{S}\}$, we rewrite  the second term in eqn.~(\ref{eqn:KL}) as a Monte Carlo integral: 
\begin{equation}
    \mathrm{loss}_\mathrm{NN} (\boldsymbol{\phi}\mid \{\mathbf{x},\mathbf{c},\textbf{S}\}) = - \frac{1}{N} \sum_{i=1}^{N} \ln \hat p(\mathbf{x}_i \mid  \boldsymbol{\phi} ; \mathbf{c}_i; \textbf{S}_i)
    \label{eqn:loss},
\end{equation}
where $N$ is the sample size. Minimizing the loss in eqn.~(\ref{eqn:loss}) is equivalent to finding the parameters $\boldsymbol{\phi}$ that maximizes the probability of the samples $\{\mathbf{x}\}$ given the corresponding conditionals $\{\mathbf{c}\}$, and noise covariances $\{\mathbf{S}\}$.

To evaluate the probability of a noisy sample $\mathbf{x}_i$, we need to convolve the GMM with the noise probability distribution. Assuming the noise $\boldsymbol{\epsilon}$ has a Gaussian distribution $\mathcal{N}(\boldsymbol{\epsilon}|\mathbf{0},\mathbf{S}_i)$, the convolution is trivial, and is simply the sum of $\mathbf{S}_i$ and every covariance $\hat{\mathbf{V}}_j$, due to the close of the Gaussian distribution under convolutions. The model probability can thus be evaluated via
\begin{equation}
    \hat p(\mathbf{x}_i\mid \boldsymbol{\phi};\mathbf{c}_i;\textbf{S}_i) = \sum_{j=1}^K \hat{\boldsymbol{\alpha}}_j(\boldsymbol{\phi};\mathbf{c}_i) \mathcal{N}\left(\mathbf{x}_i\mid \hat{\boldsymbol{\mu}}_j(\boldsymbol{\phi};\mathbf{c}_i), \hat{\mathbf{V}}_j(\boldsymbol{\phi};\mathbf{c}_i)+\textbf{S}_i\right),
    \label{eqn:noisy_GMM}
\end{equation}
at any noisy sample given its location $\mathbf{x}_i$, conditional $\mathbf{c}_i$, and noise covariance $\mathbf{S}_i$, where $\hat{\boldsymbol{\alpha}}_j$, $\hat{\boldsymbol{\mu}}_j$, and $\hat{\mathbf{V}}_j$ are the
mixing coefficients, mean, and covariance of the $j$th Gaussian.

An issue that can arise during optimization of the loss in  eqn.~(\ref{eqn:loss}) is that a Gaussian in the mixture can approach a delta function centered on a single sample. This yields an extremely large log-probability
(extremely small loss) which can eventually result in numerical overflow. As this behavior is clearly undesirable and does not represent a viable optimum, we regularize the loss by adding an additional term that amounts to a penalty when the covariance diagonal elements approaches zero:
\begin{equation}
    \mathrm{loss}_\mathrm{reg} = w\sum_j\sum_i\frac{1}{\mathrm{diag}(\mathbf{V}_j)_{i}},
    \label{eqn:loss_reg}
\end{equation}
where $w$ is a tunable parameter that we fix to $w=10^{-6}$ which we arrived at via trial and error, and $\mathrm{diag}(\mathbf{V}_j)_{i}$ is the $i$th diagonal element of the $j$th covariance. As we have forced all covariance diagonals to be positive, this regularization loss is also always positive but dominates only if the diagonal elements approach zero. The total loss is then the sum of the regularization loss and model loss
\begin{equation}
\mathrm{loss} = \mathrm{loss}_\mathrm{NN} + \mathrm{loss}_\mathrm{reg}
\end{equation}
which is what we minimize.

\subsection{Training Strategies}
\label{subsec:strategy}
We use stochastic gradient descent \citep{SDG1,SDG2} with the \texttt{torch.optim.Adam} optimizer \citep{Adam} in the \texttt{Pytorch} Python package \citep{pytorch} to train our NN. The challenges in training a neural network are to choose the proper learning rate and prevent overfitting.

We implement two methods to avoid overfitting. First, we utilize the weight decay method that introduces an additional loss term accounting for the sum of squares of the NN weights, with a coefficient of $0.001$ in the \texttt{Adam} optimizer. It penalizes large NN weight values and encourages some weights to be close to 0, i.e. to prefer a simple model. Second, the $\{\mathbf{x}, \mathbf{c}, \textbf{S}\}$ triplets are also randomly split into two sets: a training set and a validation set with ratio $90\%:10\%$. 
As long as the validation loss remains close to the training loss the model is not overfitting the training set. In each set the samples are further divided into mini-batches with size equal to $250$ samples. This number is determined rather randomly at a typical value in the literature. Compared with the sample size of 90,000 in our toy model in \S \ref{sec:exp} and 1,902,071 in the quasar contaminants in \S \ref{sec:app}, the mini-batch is still a small size. It can be increased as long as one whole mini-batch still fits in the computer memory. Stochastic gradient descent is performed by executing optimization steps based on the loss computed on each mini-batch of 250.  After looping over all of the mini-batches, we compute the average training loss of the whole training set. We then compute the validation loss, which is the loss averaged over the entire validation set.  An epoch is defined to be the execution of stochastic gradient descent on all the mini-batches plus the computation of the validation loss.  The best model is defined to be that which achieves the lowest value of the validation loss after $100$ epochs.  

The learning rate is the step size by which $\boldsymbol{\phi}$ are adjusted when trained on each mini-batch. At early stages, the learning rate should be large to speed up convergence, while, later, it should be small to allow $\boldsymbol{\phi}$ converge on precise values. The \texttt{Adam} optimizer automatically decreases the learning rate, while we implement an additional decrease. We set the initial learning rate to $0.001$ in the \texttt{Adam} optimizer, and decrease it further by multiplying $0.4$ every time when there is no decrease of the validation loss for two subsequent epochs. The latter is achieved with the \texttt{torch.optim.lr\_scheduler.ReduceLROnPlateau} scheduler.

\section{Experiments on a Simulated Noisy GMM}
\label{sec:exp}

\subsection{Constructing the GMM Toy Model}
\label{sec:toy_model}
To test the performance of our CondXD method when estimating the underlying density given observations with heteroscedastic noise, we constructed a simple toy model using a GMM with $K=10$ Gaussian components and $D=7$ dimensions. To construct the model, we first generate the mixing coefficients $\boldsymbol{\alpha}$, means $\boldsymbol{\mu}$, and Cholesky factors $\mathbf{L}$ of the covariances as a function of the conditional. However, for simplicity we only consider the case of a 1-D conditional $c$, although our method can be generalized to an $N$ dimensional conditional. 

The mixing coefficients vector $\boldsymbol{\alpha}$ is calculated using power-law functions and are integral to the generation of the Gaussian mixture. Specifically, each component of the Gaussian mixture's mixing coefficients, $\alpha_i$, is computed as:
\begin{equation}
\begin{aligned}
    &\alpha_{i,0} = A^{1-i/10} c^{1+i/10}\\
    &\alpha_{i} = \frac{\alpha_{i,0}}{\sum_i\alpha_{i,0}},
\end{aligned}
\label{eqn:exp_mixingcoeff}
\end{equation}
where $A$ is a number drawn from the uniform distribution in the range $[0,\ 2]$\footnote{These were actually generated by permuting random integers and are hence constrained to be integer multiples of $0.02$.}. This sequence introduces sufficient randomness into the mixing coefficients calculation process while constraining the range of values. The formulation of the mixing coefficients ensures that each $\alpha_i$ varies distinctively with the conditional while collectively summing to unity.

The means for our Gaussian components are generated similarly. We randomly draw $K\times D$ numbers from the uniform distribution in the range $[0,\ 10]$\footnote{Randomly sampling and permuting non-repeating integers in $[0,\ 10 \times K \times D]$ and then multiplying with $1/(K\times D)$.}. The $K\times D$ numbers are reshaped into a matrix $\mathbf{B}$ with shape $(K,\ D)$, and the means are computed as:
\begin{equation}
    \boldsymbol{\mu} = (\mathbf{B}-\overline{\mathbf{B}})\cdot c^{1.2},
    \label{eqn:exp_mean}
\end{equation}
where $\overline{\mathbf{B}}$ denotes the average of all the elements in $\mathbf{B}$ over both dimensions. For simplicity we keep using a power-law behavior on the conditional, and the exponent $1.2$ is randomly chosen and is different from that of the mixing coefficients. By subtracting $\overline{\mathbf{B}}$ from $\mathbf{B}$ we effectively center the elements of the means such that the Gaussian clusters will be evenly distributed about the origin, which simplifies the training of the NN.

The generation of Cholesky factors follows a slightly different process. We opt to generate the diagonal and off-diagonal elements respectively. We first retrieve $K\times D$ random numbers from the uniform distribution in the range $[0,\ 0.2]$\footnote{Randomly sampling and permuting non-repeating integers in $[0,\ 10 \times K \times D]$ and multiplying with $1/(50 \times K \times D)$.}. Then these numbers are reshaped into an array $\mathbf{C}_1$ of dimensions $(K, D)$. Simultaneously, we randomly select $K\times D\times(D-1)//2$ numbers from the uniform distribution in the range $[0,\ 0.2]$\footnote{Randomly sampling and permuting non-repeating integers in $[0, 10\times K \times D \times (D-1)//2]$ and multiplying with $1/(50\times K \times D \times (D-1)//2)$.}. These numbers are then reshaped into an array $\mathbf{C}_2$ of dimensions $(K, D\times(D-1)//2)$. Finally, we compute the Cholesky factor $\mathbf{L}$ as follows:
\begin{equation}
\begin{aligned}
    &\mathbf{L}_\mathrm{d} = \mathbf{C}_1 \cdot c^{0.5} + 0.1^{0.5},\\ 
    &\mathbf{L}_\mathrm{l} = \mathbf{C}_2 \cdot c^{0.5},
\end{aligned}
    \label{eqn:exp_chole}
\end{equation}
where $\mathbf{L}_{\mathrm{d}}$ represents the diagonal part, and $\mathbf{L}_{\mathrm{l}}$ represents the unique off-diagonal elements of the lower diagonal Cholesky factor $\mathbf{L}$. To ensure the positive definiteness of the covariances $\mathbf{V}$, a small constant factor of $0.1^{0.5}$ is added to $\mathbf{L}_\mathrm{d}$, which guarantees that the diagonal elements of $\mathbf{V}$ are always greater than 0.1. The exponents $0.5$ on the conditionals allows $\mathbf{C}_1$ and $\mathbf{C}_2$ to intuitively indicate the level of covariance, instead of having to intuit them from the Cholesky factor. With the Cholesky factors $\mathbf{L}$ we can compute the underlying noiseless covariance of the toy model as:
\begin{equation}
    \mathbf{V} = \mathbf{L}\mathbf{L}^\mathrm{T}.
    \label{eqn:exp_covar}
\end{equation}

In practice, real-world samples are always subject to noise. To construct a noisy toy model, we introduce the noise covariance matrices $\mathbf{S}$ using:
\begin{equation}
    \mathbf{S} = \mathbf{L}_\mathbf{S}\mathbf{L_S}^\mathrm{T}.
\label{eqn:exp_noise}
\end{equation}
In this equation, the Cholesky factor $\mathbf{L}_\mathbf{S}$ is responsible for modeling the noise characteristics. The diagonal part of the $\mathbf{L_S}$ is sampled from a uniform distribution $U(0, 1)$, while the lower-left part is sampled from another uniform distribution $U(-0.5, 0.5)$. This choice of distribution introduces both positive and negative elements in the noise covariance Cholesky factors, simulating non-trivial covariant noise in a real astronomical application. Following eqn. (\ref{eqn:noisy_GMM}), the noise covariance $\mathbf{S}$ can be added to the underlying covariance $\mathbf{V}$ in eqn. (\ref{eqn:exp_covar}) to obtain the noisy distribution.

\begin{figure*}
    \includegraphics[width=\textwidth]{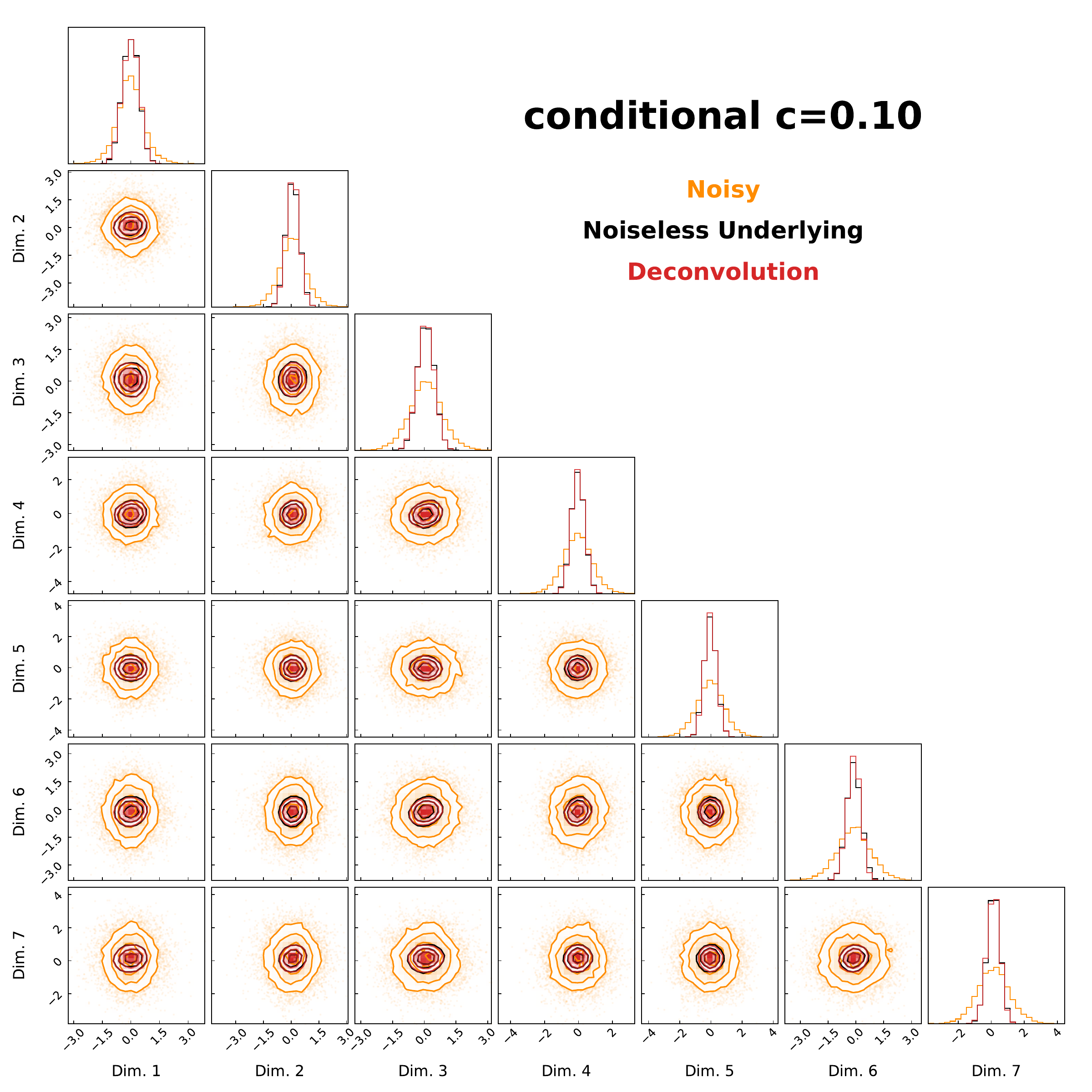}
    \caption{The distribution and density contours of $10,000$ samples from each of the noisy toy model, underlying toy model and the deconvolution when $c=0.10$. Orange scatters are samples from the noise convolved underlying distribution with orange contours representing their density contours. Black scatters and contours are for the samples from the underlying distribution, while red is for the deconvolution result. In the upper or right panels show the 1-D marginal distribution of the samples. Orange histograms represent the samples from the noise convolved underlying distribution, black is for the underlying distribution, and red is for the deconvolution. All corner plots in this paper are created by the Python package \texttt{corner} \citep{corner}.}
    \label{fig:Comp_0.1}
\end{figure*}

\begin{figure*}
    \includegraphics[width=\textwidth]{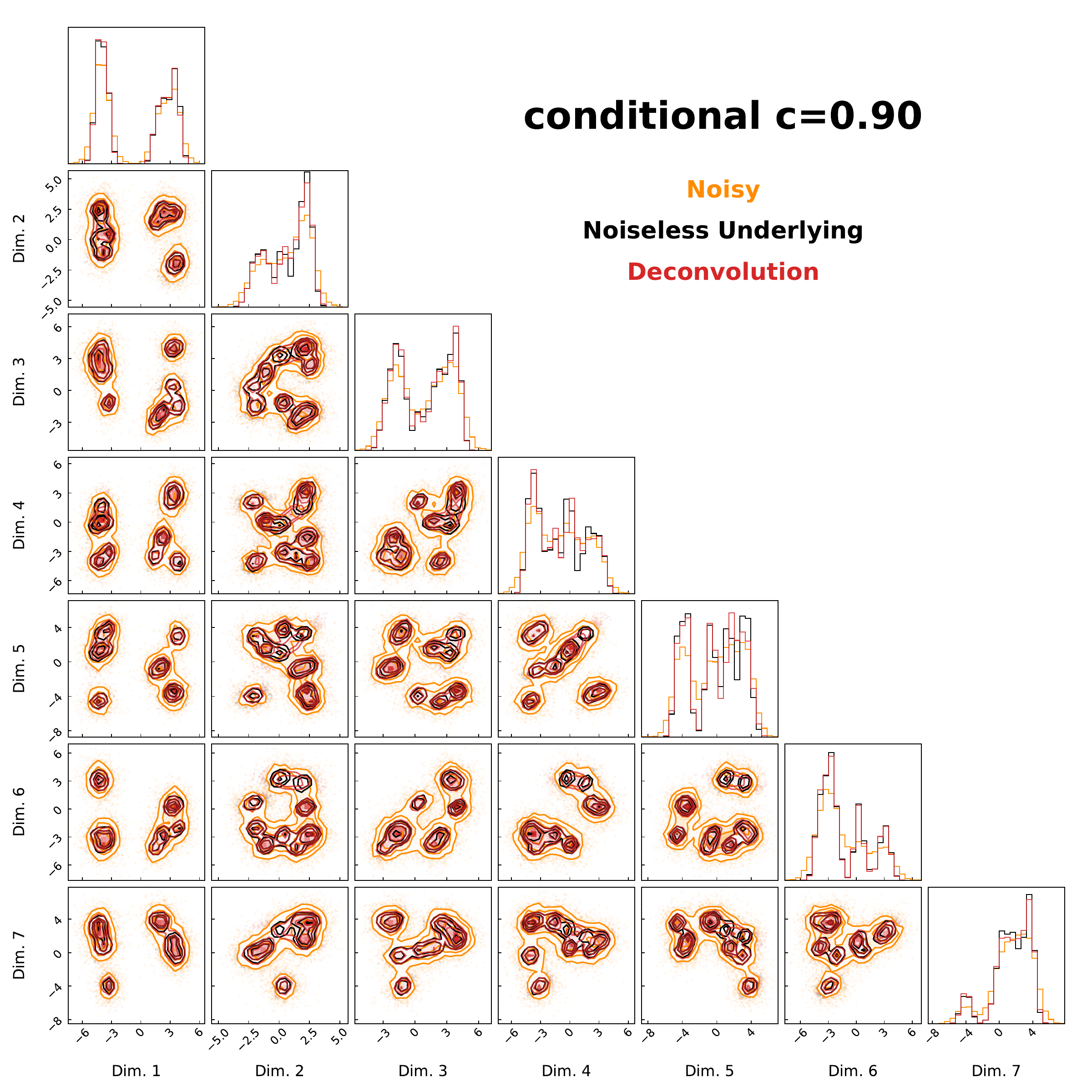}
    \caption{The distribution and density contours of $10,000$ samples from each of the noisy toy model, the underlying toy model and the deconvolution when $\mathbf{c}=0.90$. On the upper or right panels show the 1-D marginal distribution of the samples. Color scheme is the same as Fig \ref{fig:Comp_0.1}.}
    \label{fig:Comp_0.9}
\end{figure*}

The choice of power-law behavior on the conditional and the exponents used in our equations allows a broad range of behaviors for our toy model. In particular, the exponent on the conditional used in the means (eqn. \ref{eqn:exp_mean}) is larger than the one in the underlying covariance (eqn. \ref{eqn:exp_chole}). When $c$ takes on smaller values, the larger exponent in eqn. (\ref{eqn:exp_mean}) causes the Gaussian clusters to overlap. The orange points and contours in Figure \ref{fig:Comp_0.1} illustrate the samples and their densities from the noise convolved underlying distribution, in contrast to the noise free underlying distribution shown in black. The influence of our noise dominates the dispersion within the clusters. Conversely, when $c$ assumes larger values, as shown in Figure \ref{fig:Comp_0.9}, the Gaussian cluster centers separate more distinctly. Under such conditions, the influence of the noise diminishes, allowing the underlying covariances of the GMM to become more evident. 

\subsection{Training CondXD}
To illustrate the capabilities of CondXD, we generate $90,000$ $\{c, \mathbf{x}, \mathbf{S}\}$ training samples and $10,000$ validation samples from a noisy toy model defined by the simulated parameters in eqn. (\ref{eqn:exp_mixingcoeff})-(\ref{eqn:exp_covar}). To obtain a single noisy sample $\mathbf{x}$, the conditional $c$ is uniformly sampled in the range $[0,~1]$, and input in our toy model. Then, we compute the noise covariance $\mathbf{S}$ using eqn. (\ref{eqn:exp_noise}) and add it to the noiseless covariance $\mathbf{V}$, and finally draw samples $\mathbf{x}$ from this noisy distribution. Samples from the Gaussian mixture are drawn following the standard approach \citep{numpy}: the specific Gaussian cluster to be sampled is first decided via a random draw employing the the mixing coefficients as weights, and then a sample is drawn from that Gaussian cluster.

\begin{figure}
    \centering
    \includegraphics[width=\columnwidth]{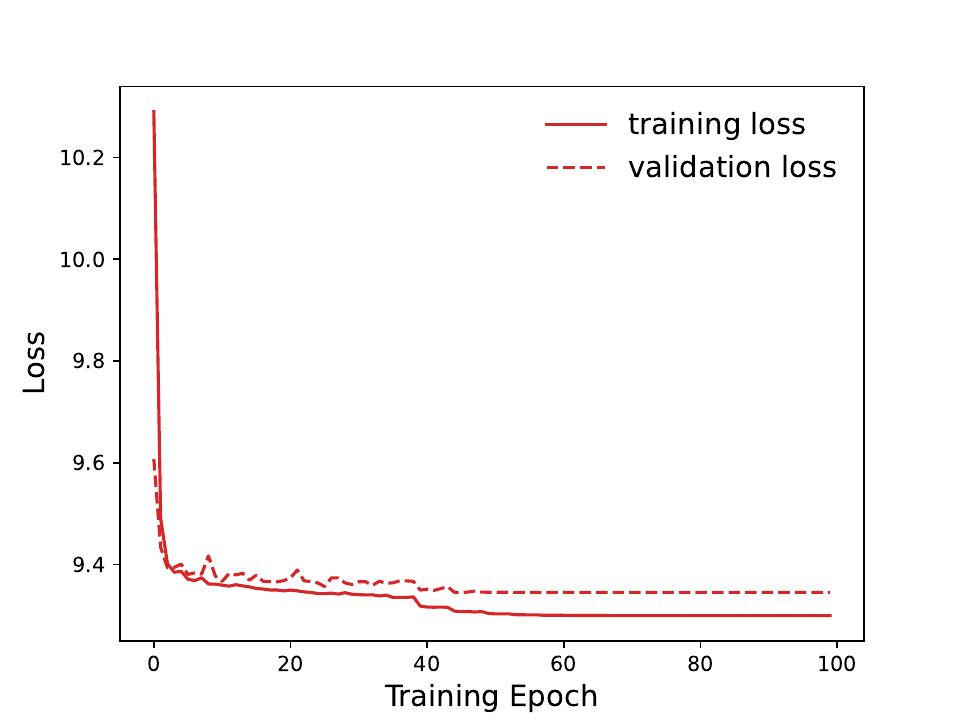}
    \caption{The loss reduction process using the 60, 000 training samples from our GMM noisy toy model. The solid red line is the training loss and dashed one is the validation loss.}
    \label{fig:loss}
\end{figure}

We train CondXD on the $90,000$ training samples with the strategies described in \S \ref{subsec:strategy}, implementing a mini-batch size of 250. After training for $100$ epochs the loss (see eqn. \ref{eqn:loss}) for the training and validation sets converge to a constant value. The training and validation loss as a function of training epoch is shown in Figure \ref{fig:loss}. Both losses decrease with training epoch, indicating that the NN has learned to fit the parameters governing the conditioned noisy distribution. In fact, overfitting is not significant, as there is only minimal disparity between the validation loss and the training loss.

\subsection{Results from the Toy Model}

In this subsection we provide a visual comparison of the aforementioned distributions. After training, to test the deconvolution capability of CondXD, we compare the estimated deconvolved distribution with the noiseless underlying distribution. Note that the underlying model is 
conditioned and we train on a continuous range of $c$ over $[0,\ 1]$, but in this section we evaluate the performance of our method only for two extreme values, $c=0.1$ and $c=0.9$, whereas the result for another intermediate case $c=0.5$ is shown in the Appendix. 

The best way to visualize how well we are deconvolving is to compare the distribution of samples drawn from our underlying noiseless model, to the distribution of samples from the trained CondXD distribution, which is usually achieved by making density contour plots of these samples. We input the specific aforementioned values of $c$ into our toy model, and generate $10,000$ samples from the underlying noiseless GMM as the test set. For comparisons with the noisy distribution, we also generate $10,000$ noisy samples by drawing $10,000$ random noise covariances, adding each to the covariance matrice of the same noiseless GMM, and sampling the noisy GMM. For CondXD, the same $c$ is input in the trained model, and $10,000$ noiseless samples are drawn from it. The density contours and 1-D marginal histograms for these three sets of samples are shown in Figure \ref{fig:Comp_0.1} ($c=0.1$) and Figure \ref{fig:Comp_0.9} ($c=0.9$).

The black histograms and contours in Figure \ref{fig:Comp_0.1} show that for $c=0.1$ the underlying Gaussians in the Gaussian mixture strongly overlap. The orange lines show the density distribution of the noisy samples from the noise convolved Gaussian mixture, which are significantly broader than the width of the underlying distribution, indicating the noise level is larger than the underlying dispersion of the Gaussian mixture. Nevertheless, CondXD still successfully deconvolves and uncovers a robust estimate of the underlying distribution. Our estimate for the deconvolved distribution is shown by the red lines. One sees qualitatively that they differ negligibly from the underlying distribution in black. 

Increasing the value of $c$ to $0.9$, the means of Gaussians separate more, as shown in Figure \ref{fig:Comp_0.9}. The orange contours of noisy samples from the GMM toy model show that the noise level is comparable to the intrinsic dispersion of the Gaussians in the mixture, which blurs the distinction between the individual components of the mixture. CondXD is still capable of estimating the noiseless underlying distribution under such conditions. Most of the red contours are consistent with the black ones, indicating that most of the individual Gaussian clusters have been recovered correctly. This is also confirmed in the panels showing the 1-D marginal distributions, as the estimated 1-D histograms differ very little from the underlying distribution. Nevertheless, in rare cases the deconvolution does not perform well. For example, in the subpanel showing dimension 4 and 6, two nearby noiseless Gaussians (black contours) are fitted with a single deconvolved Gaussian (red contours). We repeat the whole training and testing process for 10 different toy models (each has a different random seed $\xi$), and our visual assessment yields that four among the 10 realizations fail to recover all the underlying Gaussians, while the other six succeed to recover every Gaussian. In our toy model, the value of $c$ controls the separation of Gaussians. At $c=0.9$, we have almost reached the most extreme value for $c$. However, the Gaussian clusters remain insufficiently separated because the noise level amplitude is still relatively significant. As a result, CondXD struggles to perfectly differentiate every Gaussian. If we had allowed $c$ to be beyond $1$ and included more training samples, the Gaussian cluster could be more separated, and CondXD might be able to distinguish them.

\section{Comparison with Binning Method}
\label{sec:bin}
One of the main advantages introduced by the method we described in \S \ref{sec:method}, is that it can deconvolve and fit distributions that depend on conditionals. This is usually a common situation in astrophysics, where often physical properties of sources depend on other properties (e.g., the variation of the
color distributions with the magnitude of the sources). Capturing these dependencies is not an easy task, and has no standard approach. Previous works usually divide the samples into bins of conditionals, and estimate the distribution of samples in every bin respectively \citep[e.g.][]{Bovy2011, Nanni2022}. The main drawback of the binning method is that the continuity of the distribution variation, which is dependent on the conditional, among the different bins is not easily guaranteed. The distribution is supposed to vary smoothly among the bins, but the independent estimations within each bin might be trapped in some local optima, resulting in discontinuity. Furthermore, to limit the variance within each bin, the bin width should be narrow enough. However, the number of samples in each bin decreases as the bin width decreases, affecting the accuracy of the estimation. Therefore, the manual choice of a trade off between the bin width and sample size is inevitable, and there is no objective way to define it. In contrast, and the neural network of CondXD trained by all the samples naturally provides continuity, and this does not require any binning of the conditional. To demonstrate the advantages of CondXD compared to the aforementioned binning approach, we apply a binning deconvolution algorithm (denoted as bin-XD hereafter) to the GMM toy model described in \S \ref{sec:exp} and compare the results with CondXD.

Using the same training samples described \S \ref{sec:exp}, the conditionals and corresponding data samples are split into $10$ conditional bins with equal size $0.1$. Since the bins are narrow, we assume that the dependence of the sample properties with respect to the conditional inside each bin is negligible. For every bin, we apply the XDGMM method \citep{Holoien2017}, which is an implementation of the extreme deconvolution, to the training sets. XDGMM is a Python package that models mixed Gaussians with the scikit-learn API\footnote{\faGithub\ \url{https://github.com/tholoien/XDGMM}}. It performs density estimation of noisy, heterogeneous, and incomplete data with the extreme deconvolution algorithm \citep{Bovy2011XD} when an uncertainty covariance is provided, as is in our case. The hyperparameters in XDGMM are the number of Gaussians, set to $K=10$, and dimensions $D=7$, which are consistent with those used in CondXD. 
The bin-XD is progressively applied starting from the smallest conditional values ($c\in[0,~0.1]$) to the largest ones ($c\in[0.9,~1]$). The fitting in individual bins does not necessarily guarantee the continuity of the model among different bins. Following \cite{Nanni2022}, the bin-XD code fits for all the conditional bins is initialized using the best-fit parameters for the previous bin. The starting bin is the only one that is initialized without reference.

After training bin-XD, to derive the test set, we uniformly sample $25,000$ conditionals in the range $[0,~1]$, and draw $25,000$ corresponding samples from the noiseless toy model (underlying GMM). The test set is divided into the conditional bins as described in the previous paragraph. To quantify the performance of CondXD and bin-XD, we use the discrete KL divergence as a measure of the difference between the underlying density and the estimated. Similar to eqn. \ref{eqn:KL}, the discrete KL divergence is defined as:
\begin{equation}
    D_{\mathrm{KL}}(p \| \hat p;c)=\frac{1}{N}\sum_i^N \log \left(\frac{p(\mathbf{x}_i \mid c_i)}{\hat p(\mathbf{x}_i \mid c_i)}\right),
    \label{eqn:KL_exp}
\end{equation}
where $\mathbf{x}_i$ are the test samples from the underlying density, $N$ is the sample size, $p(\mathbf{x}_i\mid c_i)$ is the probability density of sample $\mathbf{x}_i$ under the underlying GMM, and $\hat p(\mathbf{x}_i\mid c_i)$ is the probability under the GMM estimated by either CondXD or bin-XD. In fact, eqn. (\ref{eqn:KL_exp}) is calculated for every conditional bin. When $p$ and $\hat p$ are close, $D_\mathrm{KL}$ should be close to zero. In general, the probability for the underlying distribution, $p(\mathbf{x}_i \mid c_i)$, should be higher than the probability for the estimated distribution $\hat p(\mathbf{x}_i \mid c_i)$, 
since they are being evaluated at samples  $\mathbf{x}_i$ from the underling distribution. Thus the KL divergence is generically expected to be positive. Besides, if we instead consider $p$ in eqn. (\ref{eqn:KL_exp}) to be the underlying noiseless GMM and $\hat p$ as the noise reconvolved estimated probability, eqn. (\ref{eqn:KL_exp}) is just the KL divergence of an algorithm that simply fits a Gaussian mixture to the noisy distribution without deconvolving. This situation represents the worst case (no deconvolution performed) and it yields a maximum value for $D_\mathrm{KL}$ which provides a useful reference. In fact, $D_\mathrm{KL}$ should lie within zero and the aforementioned maximum.

\begin{figure}
    \centering
    \includegraphics[width=\columnwidth]{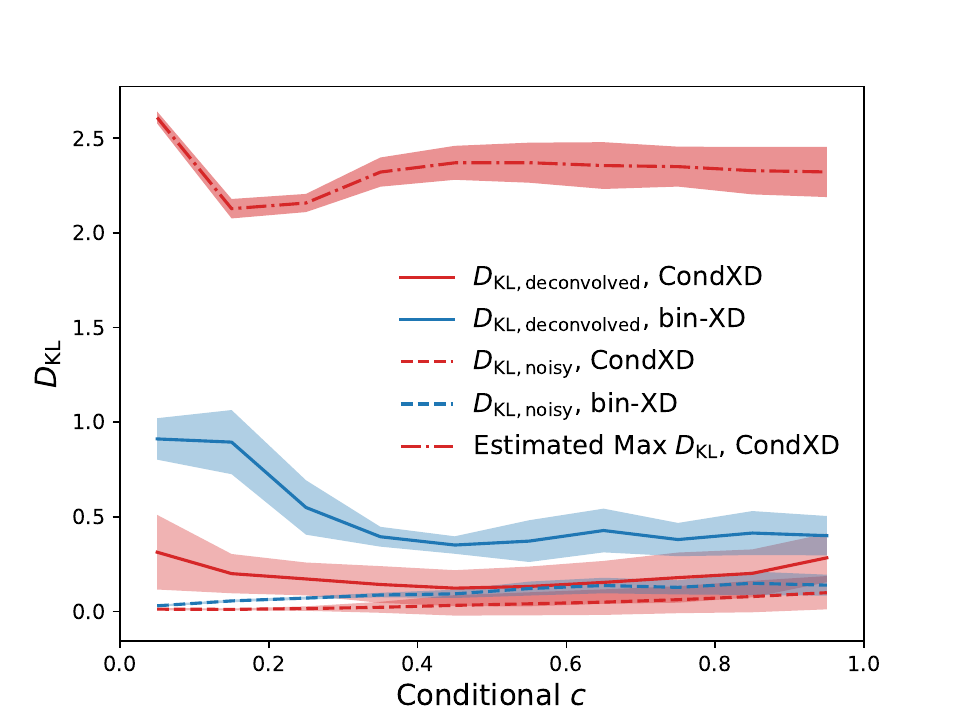}
    \caption{KL divergence of different methods as a function of the conditional $\mathbf{c}$ in our experiments. Solid lines are KL divergence measured on the underlying distributions and CondXD estimated distributions. The dashed lines are computed on the noise reconvolved underlying distributions and noise reconvolved estimated distributions. The dash-dotted line is an estimation of the possible maximum $D_\mathrm{KL}$, assuming CondXD is only fitting the noisy underlying GMM while not deconvolving at all (for details see \S \ref{sec:bin}). The red curves show the KL divergence of CondXD, while the blue curves are for bin-XD (see \S \ref{sec:bin}).}
    \label{fig:KLDiv}
\end{figure}

We compute the $D_\mathrm{KL}$ of every bin, resulting in a relation between $D_\mathrm{KL}$ and $c$. For a more general examination we repeat our experiment for ten
times with $10$ different random seeds $\xi$ that determines the toy model. In every experiment, CondXD and bin-XD are applied to the same training samples. We average the ten $D_\mathrm{KL}$ vs. $c$ curves and compute the standard deviation. The result is shown as solid curves and shaded regions respectively in Figure \ref{fig:KLDiv}. Meanwhile, we also plot the estimated maximum of $D_\mathrm{KL}$ (defined in the previous paragraph) with CondXD (dash-dotted red line in Figure \ref{fig:KLDiv}) as a reference.

Figure \ref{fig:KLDiv} shows CondXD could deconvolve (solid red line) the noisy distribution for all values of conditional $c$. The solid red line is flat and close to zero compared to the estimated maximum (dash-dotted red line). This indicates globally good performance. In contrast, bin-XD (solid blue line) shows less capability than CondXD at any conditional value, as its $D_\mathrm{KL}$ is much higher. Especially at $c\le0.2$ values, the KL divergence of the bin-XD increases remarkably. This implies that bin-XD is not a promising method for cases of overlapping Gaussians and noise domination. 

One may argue, that the poor performance of our bin-XD method at small $c$ values might be related to the fact that our fit to the lowest conditional bin is not initialized with reference to a trained bin. To verify that, we perform more experiments by training bin-XD on the opposite direction: starting from the largest $c$ bin with random initialization and proceeding toward the smallest one. However, the results are consistent with those presented in Fig \ref{fig:KLDiv} (solid blue line). The deconvolution incapability of bin-XD in the low conditional bins is intrinsic. The poor performance may result from the fact that we did not implement any strategy to prevent overfitting in the bin-XD method. As $c$ decreases and the Gaussian clusters merge, using $K=20$ Gaussians for density estimation can lead to significant degeneracy.

By evaluating the $D_\mathrm{KL}$ with $c$, we note that the value of $D_\mathrm{KL}$ of CondXD rises with the increasing of $c$. This rising of $D_\mathrm{KL}$ is likely due to the fact that CondXD is fitting two close underlying Gaussians with a single one, as described in \S \ref{sec:exp}. The noisy Gaussians in the noise convolved toy model are not separated sufficiently so that the CondXD may not be able to fit every single Gasussian correctly. If the $c$ range is broadened to larger $c$, the Gaussians are more separated, and CondXD is more likely to estimate well. 

The performance of reconstruction of the noisy distributions can also be compared if we compute a set of noise covariances from eqn. (\ref{eqn:exp_noise}) and convolve them with $p$ and $\hat p$ in eqn. (\ref{eqn:KL_exp}). The test samples $\mathbf{x}_i$ should also be re-sampled after reconvolution. We compute the same number, i.e. $25,000$, of noise covariances and draw test samples after adding the noise covariances to the underlying GMM, and calculate the KL divergence of the two noise reconvolved density distributions. The result is shown as dashed lines in Figure \ref{fig:KLDiv}. Both $D_\mathrm{KL}$ are very close to zero for all $\mathbf{c}$ values, which implies that the reconstruction is very precise. CondXD also outperforms bin-XD in the reconstruction globally.

\section{Deconvolving the Distribution of Quasar Contaminants}
\label{sec:app}
Luminous high-redshift (high-$z$) quasars are a key tool for studying the primordial universe during the epoch of reionization \citep[for recent some works see][]{Wolfson2023,Becker2021,Davies2021,Bosman2021}. However, finding the most distant quasars is challenging. Currently, only eight quasars are known at $z \geq 7$ \citep{Mortlock2011,Banados2018,Wang2018,Yang2019,Yang_2020,Matsuoka_2019_1,Matsuoka_2019_2,Wang_2021}, primarily due to the limited photometric depth of current near-infrared surveys and the decreasing number density of quasars with increasing redshift ($\approx 10^{-3} \mathrm{~deg}^{-2}$ at $J=21$, where $J$ is a flux band in the VIKING survey; \citealt{Wang2019}). Moreover, the number of contaminants, which mostly consist of cool galactic dwarfs and early-type galaxies, is much higher ($\approx 20 \mathrm{~deg}^{-2}$ at $J=21$), making efficient classification methods critical. Bayesian probabilistic methods offer a principled way to classify quasar candidates \citep[e.g.][]{Mortlock2012,Barnett2019}. One can estimate the density distribution of quasars and contaminants and compute the probability that a source belongs to quasars or contaminants (see \S \ref{sec:contaminant_estimation}).
In this section, we apply our CondXD method to a real astrophysical example: to deconvolve the flux distribution of quasars contaminants. We train our model using the same contaminant dataset described in \cite{Nanni2022}. We present the results of our deconvolution and reconstruction, as well as a brief comparison with the previous method of \cite{Nanni2022}.

\subsection{Training Data: Quasars Contaminants}
The training data we use to apply the CondXD method to the problem of high-z quasar classification is identical to the dataset described in \cite{Nanni2022}, that contains $1,902,071$ sources of quasar contaminants. In summary, our model is trained on $1076 \mathrm{~deg}^2$ of overlapping area from the DELS \citep{Dey2019}, VIKING \citep{viking2013}, and unWISE \citep{Meisner2019, Schlafly2019} imaging survey. The multi-band fluxes are obtained from DELS $z$ optical band, VIKING $YJHK_s$ near infrared (NIR) bands, and unWISE $W1W2$ mid-infrared (MIR) bands with forced photometry. The construction algorithms are described in detail in section 3.1 of \cite{Nanni2022}. The aim of \cite{Nanni2022} is to find high redshift quasars ($6 \leq z \leq 8$), whose Ly$\alpha$ lines shift to the $Y$-band, while the VIKING $J$-band could reach a depth of 22.1 at 5$\sigma$ level. Therefore, all sources in the sample are selected with high signal-to-noise ratio in $J$-band: SNR$(J)\geq5$.

\subsection{Density in the Bayesian Theorem}
\label{sec:contaminant_estimation}
To classify sources based on observed fluxes $\{\hat F\}$, we need to calculate the conditioned probability that a source belongs to a certain class according to Bayes theorem:
\begin{equation}
    P\left(O \in B \mid\left\{\hat{F}_{i}\right\}\right)=\frac{p\left(\left\{\hat{F}_{i}\right\} \mid O \in B\right) P(O \in B)}{p\left(\left\{\hat{F}_{i}\right\}\right)},
    \label{eqn:bayesian}
\end{equation}
where $O$ is the object and $B$ is the class, i.e. quasars or contaminants. If we denote quasars as $A$ and contaminants as $B$, the denominator of the right-hand side in eqn. (\ref{eqn:bayesian}) is defined as 
\begin{equation}
    p\left(\left\{\hat{F}_{i}\right\}\right)=p\left(\left\{\hat{F}_{i}\right\} \mid O \in A\right) P(O \in A)+p\left(\left\{\hat{F}_{i}\right\} \mid O \in B\right) P(O \in B),
\end{equation}
as a source can only be a quasar or contaminant. The factor $P(O \in B)$ in the numerator of the right-hand side of eqn. (\ref{eqn:bayesian}) is the prior, which could be approximated as the fraction of quasars in the data set. The other factor, $p\left(\left\{\hat{F}_{i}\right\} \mid O \in B\right)$, is the density of quasars in flux space that is to be estimated. 

The distribution functions of quasar fluxes in astronomical surveys are well described by power-law functions of their apparent magnitude, quasar luminosity functions and object number count distributions typically follow power-laws. In the context of a Gaussian mixture model, it would require a large number of Gaussian components to model this distribution accurately. In contrast, their color (logarithm of relative flux) distribution is flat enough to be modeled by a small number of Gaussians. Furthermore, the crucial information for distinguishing quasars and contaminants lies mostly in the color. This motivates people to use a distribution model based on the color. Additionally, relative fluxes are easier to derive and more straight forward to model than colors. In the case of faint sources that drop out in certain bands (e.g., high-$z$ quasars), the measured fluxes could be non-positive, and it is infeasible compute the color, i.e. logarithm of zero or a negative value. Furthermore, the observational uncertainties of the relative fluxes are closer to Gaussian than colors, especially when the uncertainty in the reference $J$-band is small. In fact, as both the numerators and denominators are noisy, the Gaussian approximation of the flux ratio density can only be validated when the noise of the denominators is small. If the noise of the the denominators is large, the distribution of the ratio of two Gaussian random variables is not Gaussian. In our case, since the observed $J$-band flux, $\hat F_J$, is always significantly detected at great than $5\sigma$ significance, this condition is well satisfied. Hence, instead of fitting the distribution of the measured fluxes, we choose to model the fluxes relative to the $J$-band flux. 

We separate the flux relative to $J$-band from the absolute flux in the likelihood as follows:
\begin{equation}
    \begin{array}{r}
        p\left(\left\{\hat F_{i}\right\} \mid O \in \text { "cont." }\right)=p\left(\left\{\hat F_{i} / \hat F_{J}\right\} \mid \hat F_{J}, O \in \text { "cont." }\right) \\
        \qquad \times p\left(\hat F_{J} \mid O \in \text { "cont." }\right),
    \end{array}
    \label{eqn:relative}
\end{equation}
where $\hat F_i$ are the fluxes of $z$, $Y$, $H$, $K$, $W1$, $W2$ bands. In this equation, the probability density of the absolute fluxes is separated into the distribution of the relative fluxes conditioned on the $J$-band flux and the distribution of the $J$-band fluxes. In this paper we mainly discuss the first factor, and the derivation of the second factor can be found in \S 4.3 in \cite{Nanni2022}.

\subsection{Density Estimation}
\cite{Nanni2022} employed their XD based XDHZQSO algorithm to fit the distribution of the contaminants and simulated high redshift quasars with a GMM. The algorithm has demonstrated high efficiency, accuracy, and stability. XDHZQSO, however, has to divide the contaminants into a discrete number of $J$-band magnitude bins, because the contaminant color distribution is a strong function of magnitude, but XDHZQSO cannot be used to estimate in the continuous limit. They implemented complicated strategies to capture the variation of the relative flux distribution with magnitude and guarantee continuity. The authors used $50$ overlapping bins, with the width of each bin determined by a broken sigmoid function of the $J$-bin right edge. As the right edges are uniformly distributed, the bins overlap with their neighbors. The overlap between the bins guarantees a continuity among adjacent bins as well as a sufficient number of sources at the faint and bright end of the $J$-band magnitude. Within each bin, they used the XD algorithm to estimate the density, with the same initialization strategy described in \S \ref{sec:bin}, to improve the model's continuity. These strategies make the training process slow, as some samples belong to multiple bins and will be input to the training process multiple times. Furthermore, this binning strategy results in additional problems, since the bin width is very large at the two ends, e.g. the resulting magnitude range of $5$ mag compared with the right edge step $0.05$ magnitude at the faintest end. This makes it hard to correctly capture the variation of the model.

\begin{figure}
    \centering
    \includegraphics[width=\columnwidth]{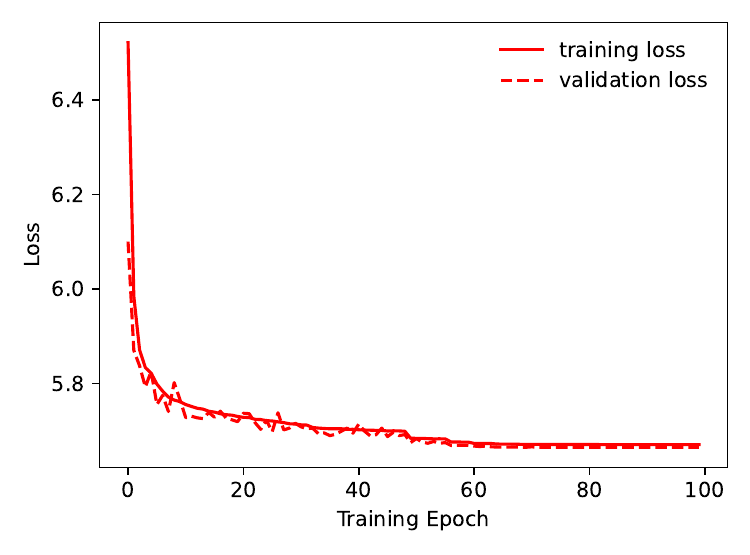}
    \caption{The loss decrease process of CondXD in the experiment of application to the quasar contaminants of \protect\cite{Nanni2022}. The solid red line is the training loss and the dashed line is the validation loss.}
    \label{fig:loss_QSO}
\end{figure}

Instead, with our CondXD method, we can treat the $J$-band magnitude as a conditional $c$ and build one continuous and general model by deriving the Gaussian parameters from the NN. We model the six-dimensional density distribution of relative fluxes $\{f_z/f_J,\ f_Y/f_J,\ f_H/f_J,\ f_{K_s}/f_J,\ f_{W1}/f_J,\ f_{W2}/f_J\}$ using $K=20$ Gaussian components. The number of Gaussians adopted is consistent with the number chosen by \cite{Bovy2011} and \cite{Nanni2022}. Empirically, models with less than $20$ components overly smooth the observed distribution, while more than $20$ components are likely to suffer from overfitting. As we are deconvolving the relative noisy fluxes instead of the measured fluxes, the uncertainty covariance matrix should be computed. The validity and derivation of the uncertainty covariance matrix of relative fluxes have been discussed in the Appendix A of \cite{Nanni2022}. Specifically, one needs to remove the off-diagonal elements (i.e. set them to $0$) in the relative flux noise covariances when $J>21$. This is because in the limit of faint $J$-band regime, the noise becomes significant compared with the flux, and the distribution of the relative fluxes violates the Gaussian assumption as discussed earlier. As we are estimating with a GMM assuming Gaussian noise, the non-Gaussian noise should be approximated by a Gaussian. We convolve the GMM output by CondXD with the uncertainties of the relative fluxes by adding the uncertainty covariance to the GMM covariance. The samples are split into training and validation set with ratio $9:1$. Training and validating the NN with the strategies described in \S \ref{subsec:strategy} for $100$ epochs, our model converges. The loss decrease is shown in Figure \ref{fig:loss_QSO}.

\begin{figure*}
  \includegraphics[width=\textwidth]{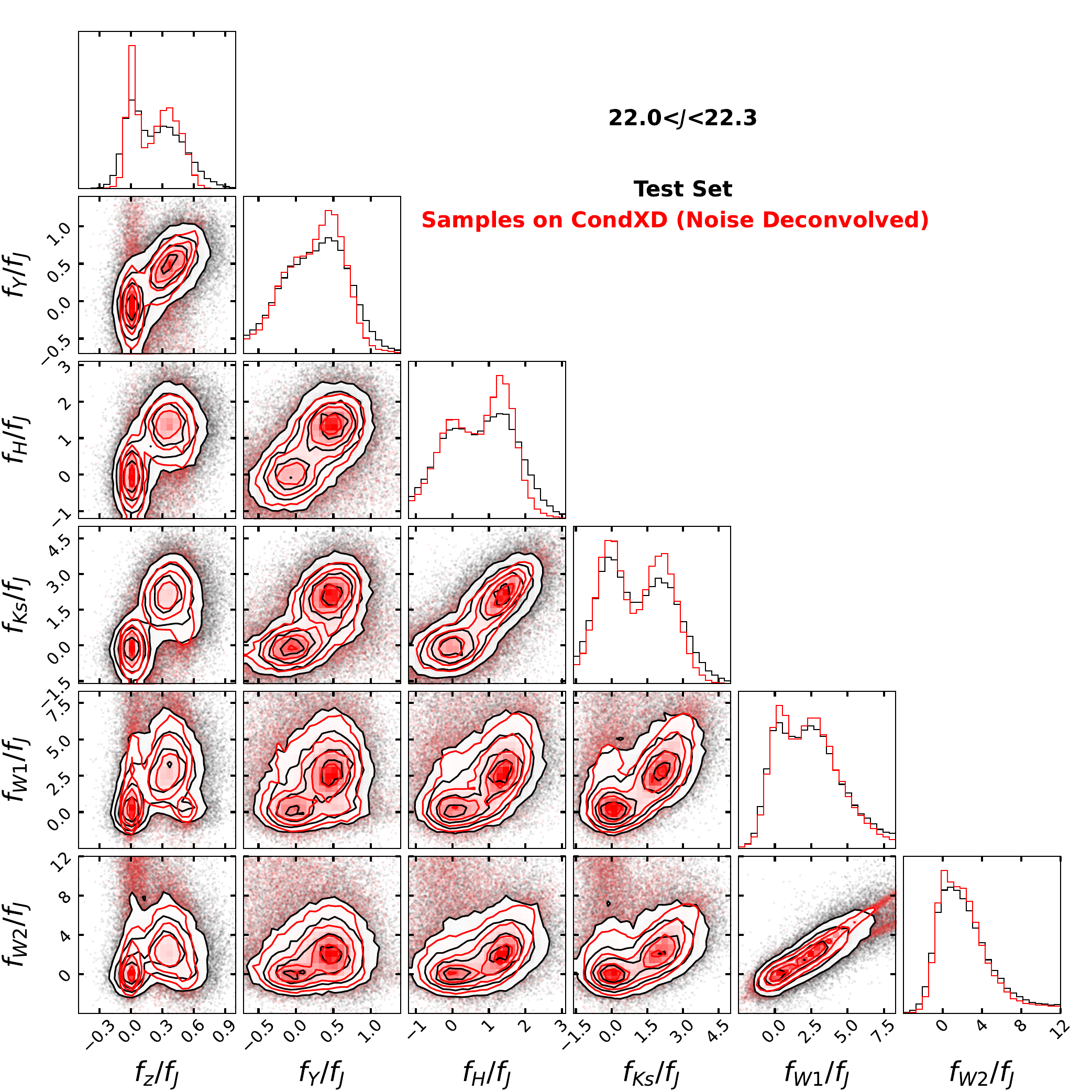}
    \caption{The relative fluxes of all quasar contaminant samples in $22.0 < J < 22.3$ bin and their density contours are plotted in black. After deconvolution with CondXD, the samples from the noise free estimation and their density contours are shown in red. The red contours are narrower because the noise has been deconvolved.}
    \label{fig:densityJbin_dec}
\end{figure*}

\begin{figure*}
  \includegraphics[width=\textwidth]{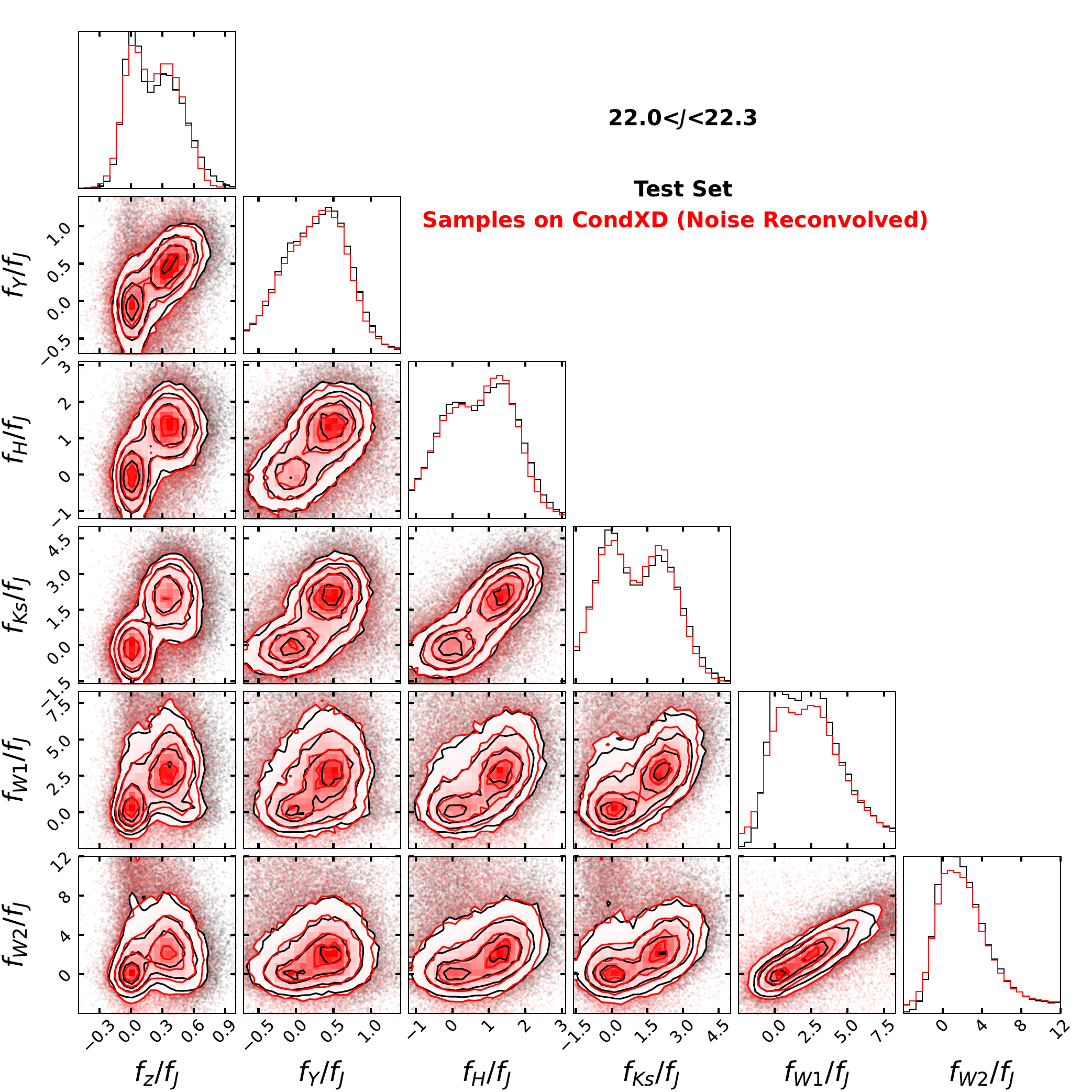}
    \caption{The relative fluxes of all quasar contaminant samples in $22.0 < J < 22.3$ bin and their density contours are plotted in black. In order to illustrate after deconvolution by CondXD we are still capable of reconstructing the noisy data, we convolve the noise free estimation in Figure \ref{fig:densityJbin} with the noise of the quasar contaminants. The samples from the noise reconvovlved model and their density contours are shown in red. There red contours has only negligible differences from the black, proving the reconstruction is successful.}
    \label{fig:densityJbin}
\end{figure*}

We compare the distribution of the whole contaminant set with the corresponding predictions by our trained model in Figure \ref{fig:densityJbin_dec} and \ref{fig:densityJbin}. As we do not have access to the underlying noiseless distribution of the relative fluxes, we can only compare our predictions, either noiseless or convolved with noise, with the noisy data set. We select the same $J$-band range as the Appendix of \cite{Nanni2022}, i.e. $22.0<J<22.3$, for display and comparison purposes. For each object in this $J$-band bin, its $J$-band magnitude is input to the CondXD model and a GMM is output. Then, for each object one noiseless predicted data point is sampled from the GMM. By convolving the GMM with the source's uncertainty distribution, we can also sample a noisy prediction. The distribution of the noisy predictions are shown in Fig \ref{fig:densityJbin}. 
Comparing Figure \ref{fig:densityJbin_dec} with Figure A1 in \cite{Nanni2022}, the two deconvolutions yield similar results. In Figure \ref{fig:densityJbin}, the noisy prediction distribution (red) matches the original samples (black) promisingly. CondXD has reconstructed the noisy distribution precisely. Compared with Figure A2 in \cite{Nanni2022}, our model performs similarly to theirs. The distributions of the noisy predictions (red) in all the other bins produced with our model are also consistent with those from \cite{Nanni2022}. Besides of performance, our model finishes training on within three hours on $1,902,071$ samples with a 2.8 GHz Quad-Core Intel Core i7 for Macbook, compared with $\approx30$ hours with their model. This is partly because many data in the overlap of different bins are used for training for multiple times, which largely increased the time required to construct a model for all the bins. Note that no GPU is implemented in any of our experiments. With a GPU the time cost can be greatly reduced.

\section{Conclusions and Discussion}
\label{sec:conclution} 
In this paper we built a conditional density deconvolution algorithm, CondXD, with a neural network. This is an extension of the existing XD method and a combination with the mixture density networks. It features in the ability to estimate the underlying density of noisy properties that depends on some conditionals, given a set of data with large and heteroscedastic uncertainties. The code is available at \url{https://github.com/enigma-igm/CondXD}.

We experiment CondXD on a toy model, a GMM whose parameters (i.e. mixing coefficients, means and covariances) are dependent on a conditional. The samples are drawn from the GMM convolved with non-identical noise covariances. The result shows CondXD is able to deconvolve the heteroscedastic uncertainties and estimate the underlying conditional dependent GMM. It can also reconstruct the noisy distribution given the noise. Further experiments performing a classic binning XD on the same toy model show that CondXD is remarkably more capable than the binning methods in both continuity and accuracy. It exhibits a flat $D_\mathrm{KL}$ curve across the conditional range, which is globally smaller than the binning method, indicating comprehensively more solid estimation. Especially in the low signal-to-noise ratio region ($c<0.05$ in Figure \ref{fig:KLDiv}), the $D_\mathrm{KL}$ of CondXD is close to $0$, while the binning method approaches the estimated worst. We further apply our method to a real astronomical case, i.e. inferring the underlying distribution of a set of noisy high-$z$ quasar contaminants fluxes. Compared with the method used by \citet{Nanni2022}, which used a binning approach, our method outputs a comparable result, but $\approx 10$ times faster.

Although we apply CondXD to only 1-D conditionals, it can be easily generalized to multi-dimensional conditional cases. For example, \cite{Bovy2012} included not only the reference band flux but also the redshift as new features in addition to the original band fluxes, in order to obtain the flux density in different redshift ranges. With our approach we don't need to add an extra feature dimension. There is no appreciable uncertainties on redshift, as quasar colors do not vary significantly within typical redshift uncertainties. Therefore, redshift perfectly matches our requirement that the noise of conditionals should be negligible. In conclusion, redshift is certainly another reasonable conditional that is worth including. 

Nonetheless, restrictions still exist in our algorithm. This method only deconvolves the features, and it cannot deconvolve the conditional. Our conditionals need to be noiseless, while this is rarely satisfied in practice, like in \S \ref{sec:app}. Therefore, the conditionals should all have high SNR to approximate the noise-free assumption. Furthermore, the target distribution should also be well approximated by a Gaussian mixture. This is also the motivation of converting fluxes to relative fluxes in our experiment. Another commonly used approach in density estimation that could possibly help to solve such issues is a normalizing flow \citep{Tabak2010, Tabak2013}. Normalizing flows transform a density which is easy to describe into a complicated density by a set of invertible functions, and have shown good scalability and flexibility in density estimation \citep[e.g.][]{Rezende2015,Cranmer2019}. This class of methods do not require any feature (dimension) of data to be noise free, nor any marginal distribution to be Gaussian. Although these works did not consider the conditional densities, normalizing flows can take the conditionals as new features (dimensions) and deconvolve the general distribution, and further compute the conditional density like in \cite{Bovy2012}. However, to our knowledge only homoscedastic noise (identical noise distribution for all samples) has been considered \citep{Dockhorn2020}. Our CondXD might still be the best method for deconvolving conditional densities with heteroscedastic noise.

\section*{Acknowledgements}
We would like to express our sincere appreciation to the constructive comments by the ENIGMA group at Leiden Observatory and UCSB on this project. We also extend our gratitude to James A. Ritchie and Iain Murray for making their code \texttt{scalable\_xd} available, which served as a foundation for parts of our own code. Additionally, we are thankful to Tom Holoien, Phil Marshall and Risa Wechsler for providing their \texttt{XDGMM} code, which made it possible to conduct our experiment comparing our model with the classic binning method. 

We are also grateful to the developers and contributors of the following Python packages, whose invaluable tools and libraries greatly facilitated our research: \texttt{Numpy} for providing efficient array operations and numerical computing capabilities that formed the backbone of our data analysis and modeling; \texttt{Astropy} for its comprehensive set of astronomical tools and utilities, enabling us to manipulate astronomical data with ease; \texttt{PyTorch} for its cutting-edge deep learning framework, which empowered us to implement and train our complex neural network model; \texttt{Matplotlib} and \texttt{corner} for its powerful visualization capabilities, which allowed us to create informative plots to present our results.





\bibliographystyle{mnras}
\bibliography{example} 

\begin{thebibliography}{}
\makeatletter
\relax
\def\mn@urlcharsother{\let\do\@makeother \do\$\do\&\do\#\do\^\do\_\do\%\do\~}
\def\mn@doi{\begingroup\mn@urlcharsother \@ifnextchar [ {\mn@doi@}
  {\mn@doi@[]}}
\def\mn@doi@[#1]#2{\def\@tempa{#1}\ifx\@tempa\@empty \href
  {http://dx.doi.org/#2} {doi:#2}\else \href {http://dx.doi.org/#2} {#1}\fi
  \endgroup}
\def\mn@eprint#1#2{\mn@eprint@#1:#2::\@nil}
\def\mn@eprint@arXiv#1{\href {http://arxiv.org/abs/#1} {{\tt arXiv:#1}}}
\def\mn@eprint@dblp#1{\href {http://dblp.uni-trier.de/rec/bibtex/#1.xml}
  {dblp:#1}}
\def\mn@eprint@#1:#2:#3:#4\@nil{\def\@tempa {#1}\def\@tempb {#2}\def\@tempc
  {#3}\ifx \@tempc \@empty \let \@tempc \@tempb \let \@tempb \@tempa \fi \ifx
  \@tempb \@empty \def\@tempb {arXiv}\fi \@ifundefined
  {mn@eprint@\@tempb}{\@tempb:\@tempc}{\expandafter \expandafter \csname
  mn@eprint@\@tempb\endcsname \expandafter{\@tempc}}}

\bibitem[\protect\citeauthoryear{{Ba{\~n}ados} et~al.,}{{Ba{\~n}ados}
  et~al.}{2018}]{Banados2018}
{Ba{\~n}ados} E.,  et~al., 2018, \mn@doi [\nat] {10.1038/nature25180}, \href
  {https://ui.adsabs.harvard.edu/abs/2018Natur.553..473B} {553, 473}

\bibitem[\protect\citeauthoryear{{Becker}, {D'Aloisio}, {Christenson}, {Zhu},
  {Worseck}  \& {Bolton}}{{Becker} et~al.}{2021}]{Becker2021}
{Becker} G.~D.,  {D'Aloisio} A.,  {Christenson} H.~M.,  {Zhu} Y.,  {Worseck}
  G.,   {Bolton} J.~S.,  2021, \mn@doi [\mnras] {10.1093/mnras/stab2696}, \href
  {https://ui.adsabs.harvard.edu/abs/2021MNRAS.508.1853B} {508, 1853}

\bibitem[\protect\citeauthoryear{{Bird}, {Xue}, {Liu}, {Shen}, {Flynn}, {Yang},
  {Zhao}  \& {Tian}}{{Bird} et~al.}{2021}]{Bird2021}
{Bird} S.~A.,  {Xue} X.-X.,  {Liu} C.,  {Shen} J.,  {Flynn} C.,  {Yang} C.,
  {Zhao} G.,   {Tian} H.-J.,  2021, \mn@doi [\apj] {10.3847/1538-4357/abfa9e},
  \href {https://ui.adsabs.harvard.edu/abs/2021ApJ...919...66B} {919, 66}

\bibitem[\protect\citeauthoryear{Bishop}{Bishop}{2006}]{Bishop}
Bishop C.~M.,  2006, Pattern Recognition and Machine Learning.
Springer New York, NY

\bibitem[\protect\citeauthoryear{{Bosman}}{{Bosman}}{2021}]{Bosman2021}
{Bosman} S. E.~I.,  2021, \mn@doi [arXiv e-prints] {10.48550/arXiv.2108.12446},
  \href {https://ui.adsabs.harvard.edu/abs/2021arXiv210812446B} {p.
  arXiv:2108.12446}

\bibitem[\protect\citeauthoryear{{Bovy}, {Hogg}  \& {Roweis}}{{Bovy}
  et~al.}{2011a}]{Bovy2011XD}
{Bovy} J.,  {Hogg} D.~W.,   {Roweis} S.~T.,  2011a, \mn@doi [Annals of Applied
  Statistics] {10.1214/10-AOAS439}, \href
  {https://ui.adsabs.harvard.edu/abs/2011AnApS...5.1657B} {5, 1657}

\bibitem[\protect\citeauthoryear{{Bovy} et~al.,}{{Bovy}
  et~al.}{2011b}]{Bovy2011}
{Bovy} J.,  et~al., 2011b, \mn@doi [\apj] {10.1088/0004-637X/729/2/141}, \href
  {https://ui.adsabs.harvard.edu/abs/2011ApJ...729..141B} {729, 141}

\bibitem[\protect\citeauthoryear{{Bovy} et~al.,}{{Bovy}
  et~al.}{2012}]{Bovy2012}
{Bovy} J.,  et~al., 2012, \mn@doi [\apj] {10.1088/0004-637X/749/1/41}, \href
  {https://ui.adsabs.harvard.edu/abs/2012ApJ...749...41B} {749, 41}

\bibitem[\protect\citeauthoryear{{Buder} et~al.,}{{Buder}
  et~al.}{2022}]{Buder2022}
{Buder} S.,  et~al., 2022, \mn@doi [\mnras] {10.1093/mnras/stab3504}, \href
  {https://ui.adsabs.harvard.edu/abs/2022MNRAS.510.2407B} {510, 2407}

\bibitem[\protect\citeauthoryear{{Cranmer}, {Galvez}, {Anderson}, {Spergel}  \&
  {Ho}}{{Cranmer} et~al.}{2019}]{Cranmer2019}
{Cranmer} M.~D.,  {Galvez} R.,  {Anderson} L.,  {Spergel} D.~N.,   {Ho} S.,
  2019, arXiv e-prints, \href
  {https://ui.adsabs.harvard.edu/abs/2019arXiv190808045C} {p. arXiv:1908.08045}

\bibitem[\protect\citeauthoryear{{Davies}, {Bosman}, {Furlanetto}, {Becker}  \&
  {D'Aloisio}}{{Davies} et~al.}{2021}]{Davies2021}
{Davies} F.~B.,  {Bosman} S. E.~I.,  {Furlanetto} S.~R.,  {Becker} G.~D.,
  {D'Aloisio} A.,  2021, \mn@doi [\apjl] {10.3847/2041-8213/ac1ffb}, \href
  {https://ui.adsabs.harvard.edu/abs/2021ApJ...918L..35D} {918, L35}

\bibitem[\protect\citeauthoryear{Devroye}{Devroye}{1989}]{Devroye1989}
Devroye L.,  1989, The Canadian Journal of Statistics / La Revue Canadienne de
  Statistique, 17, 235

\bibitem[\protect\citeauthoryear{{Dey} et~al.,}{{Dey} et~al.}{2019}]{Dey2019}
{Dey} A.,  et~al., 2019, \mn@doi [\aj] {10.3847/1538-3881/ab089d}, \href
  {https://ui.adsabs.harvard.edu/abs/2019AJ....157..168D} {157, 168}

\bibitem[\protect\citeauthoryear{DiPompeo, Bovy, Myers  \& Lang}{DiPompeo
  et~al.}{2015}]{DiPompeo2015}
DiPompeo M.~A.,  Bovy J.,  Myers A.~D.,   Lang D.,  2015, \mn@doi [Monthly
  Notices of the Royal Astronomical Society] {10.1093/mnras/stv1562}, 452, 3124

\bibitem[\protect\citeauthoryear{{Dockhorn}, {Ritchie}, {Yu}  \&
  {Murray}}{{Dockhorn} et~al.}{2020}]{Dockhorn2020}
{Dockhorn} T.,  {Ritchie} J.~A.,  {Yu} Y.,   {Murray} I.,  2020, arXiv
  e-prints, \href {https://ui.adsabs.harvard.edu/abs/2020arXiv200609396D} {p.
  arXiv:2006.09396}

\bibitem[\protect\citeauthoryear{{Edge}, {Sutherland}, {Kuijken}, {Driver},
  {McMahon}, {Eales}  \& {Emerson}}{{Edge} et~al.}{2013}]{viking2013}
{Edge} A.,  {Sutherland} W.,  {Kuijken} K.,  {Driver} S.,  {McMahon} R.,
  {Eales} S.,   {Emerson} J.~P.,  2013, The Messenger, \href
  {https://ui.adsabs.harvard.edu/abs/2013Msngr.154...32E} {154, 32}

\bibitem[\protect\citeauthoryear{{Euclid Collaboration} et~al.,}{{Euclid
  Collaboration} et~al.}{2019}]{Barnett2019}
{Euclid Collaboration} et~al., 2019, \mn@doi [A&A]
  {10.1051/0004-6361/201936427}, 631, A85

\bibitem[\protect\citeauthoryear{Fan}{Fan}{1991a}]{Fan1991a}
Fan J.,  1991a, Statistica Sinica, 1, 541

\bibitem[\protect\citeauthoryear{Fan}{Fan}{1991b}]{Fan1991b}
Fan J.,  1991b, \mn@doi [The Annals of Statistics] {10.1214/aos/1176348248},
  19, 1257

\bibitem[\protect\citeauthoryear{Foreman-Mackey}{Foreman-Mackey}{2016}]{corner}
Foreman-Mackey D.,  2016, \mn@doi [The Journal of Open Source Software]
  {10.21105/joss.00024}, 1, 24

\bibitem[\protect\citeauthoryear{{Gepperth} \& {Pf{\"u}lb}}{{Gepperth} \&
  {Pf{\"u}lb}}{2019}]{Gepperth2019}
{Gepperth} A.,  {Pf{\"u}lb} B.,  2019, arXiv e-prints, \href
  {https://ui.adsabs.harvard.edu/abs/2019arXiv191209379G} {p. arXiv:1912.09379}

\bibitem[\protect\citeauthoryear{Harris et~al.,}{Harris et~al.}{2020}]{numpy}
Harris C.~R.,  et~al., 2020, \mn@doi [Nature] {10.1038/s41586-020-2649-2}, 585,
  357

\bibitem[\protect\citeauthoryear{He, Zhang, Ren  \& Sun}{He
  et~al.}{2015}]{PReLU}
He K.,  Zhang X.,  Ren S.,   Sun J.,  2015, in 2015 IEEE International
  Conference on Computer Vision (ICCV). pp 1026--1034,
  \mn@doi{10.1109/ICCV.2015.123}

\bibitem[\protect\citeauthoryear{{Holoien}, {Marshall}  \&
  {Wechsler}}{{Holoien} et~al.}{2017}]{Holoien2017}
{Holoien} T. W.~S.,  {Marshall} P.~J.,   {Wechsler} R.~H.,  2017, \mn@doi [\aj]
  {10.3847/1538-3881/aa68a1}, \href
  {https://ui.adsabs.harvard.edu/abs/2017AJ....153..249H} {153, 249}

\bibitem[\protect\citeauthoryear{Hosseini \& Sra}{Hosseini \&
  Sra}{2015}]{Hosseini2015}
Hosseini R.,  Sra S.,  2015, Advances in Neural Information Processing Systems,
  28, 910

\bibitem[\protect\citeauthoryear{Hosseini \& Sra}{Hosseini \&
  Sra}{2020}]{Hosseini2017}
Hosseini R.,  Sra S.,  2020, \mn@doi [Mathematical Programming]
  {10.1007/s10107-019-01381-4}, 181, 187

\bibitem[\protect\citeauthoryear{{Ivezi{\'c}} \& {Ivezi{\'c}}}{{Ivezi{\'c}} \&
  {Ivezi{\'c}}}{2021}]{Ivezic2021}
{Ivezi{\'c}} V.,  {Ivezi{\'c}} {\v{Z}}.,  2021, \mn@doi [\icarus]
  {10.1016/j.icarus.2020.114262}, \href
  {https://ui.adsabs.harvard.edu/abs/2021Icar..35714262I} {357, 114262}

\bibitem[\protect\citeauthoryear{{Ivezi{\'c}} et~al.,}{{Ivezi{\'c}}
  et~al.}{2019}]{Ivezic2019}
{Ivezi{\'c}} {\v{Z}}.,  et~al., 2019, \mn@doi [\apj]
  {10.3847/1538-4357/ab042c}, \href
  {https://ui.adsabs.harvard.edu/abs/2019ApJ...873..111I} {873, 111}

\bibitem[\protect\citeauthoryear{{Jimenez Rezende} \& {Mohamed}}{{Jimenez
  Rezende} \& {Mohamed}}{2015}]{Rezende2015}
{Jimenez Rezende} D.,  {Mohamed} S.,  2015, arXiv e-prints, \href
  {https://ui.adsabs.harvard.edu/abs/2015arXiv150505770J} {p. arXiv:1505.05770}

\bibitem[\protect\citeauthoryear{Kiefer \& Wolfowitz}{Kiefer \&
  Wolfowitz}{1952}]{SDG2}
Kiefer J.,  Wolfowitz J.,  1952, \mn@doi [The Annals of Mathematical
  Statistics] {10.1214/aoms/1177729392}, 23, 462

\bibitem[\protect\citeauthoryear{Kingma \& Ba}{Kingma \& Ba}{2014}]{Adam}
Kingma D.~P.,  Ba J.,  2014, Adam: A Method for Stochastic Optimization,
  \mn@doi{10.48550/ARXIV.1412.6980}, \url {https://arxiv.org/abs/1412.6980}

\bibitem[\protect\citeauthoryear{Kullback \& Leibler}{Kullback \&
  Leibler}{1951}]{D_KL}
Kullback S.,  Leibler R.~A.,  1951, \mn@doi [The Annals of Mathematical
  Statistics] {10.1214/aoms/1177729694}, 22, 79

\bibitem[\protect\citeauthoryear{{Matsuoka} et~al.,}{{Matsuoka}
  et~al.}{2019a}]{Matsuoka_2019_1}
{Matsuoka} Y.,  et~al., 2019a, \mn@doi [\apj] {10.3847/1538-4357/ab3c60}, \href
  {https://ui.adsabs.harvard.edu/abs/2019ApJ...883..183M} {883, 183}

\bibitem[\protect\citeauthoryear{{Matsuoka} et~al.,}{{Matsuoka}
  et~al.}{2019b}]{Matsuoka_2019_2}
{Matsuoka} Y.,  et~al., 2019b, \mn@doi [\apj] {10.3847/1538-4357/ab3c60}, \href
  {https://ui.adsabs.harvard.edu/abs/2019ApJ...883..183M} {883, 183}

\bibitem[\protect\citeauthoryear{Meisner, Lang, Schlafly  \& Schlegel}{Meisner
  et~al.}{2019}]{Meisner2019}
Meisner A.~M.,  Lang D.,  Schlafly E.~F.,   Schlegel D.~J.,  2019, \mn@doi
  [Publications of the Astronomical Society of the Pacific]
  {10.1088/1538-3873/ab3df4}, 131, 124504

\bibitem[\protect\citeauthoryear{Mortlock, Patel, Warren, Hewett, Venemans,
  McMahon  \& Simpson}{Mortlock et~al.}{2011a}]{Mortlock2012}
Mortlock D.~J.,  Patel M.,  Warren S.~J.,  Hewett P.~C.,  Venemans B.~P.,
  McMahon R.~G.,   Simpson C.,  2011a, \mn@doi [Monthly Notices of the Royal
  Astronomical Society] {10.1111/j.1365-2966.2011.19710.x}, 419, 390

\bibitem[\protect\citeauthoryear{{Mortlock} et~al.,}{{Mortlock}
  et~al.}{2011b}]{Mortlock2011}
{Mortlock} D.~J.,  et~al., 2011b, \mn@doi [\nat] {10.1038/nature10159}, \href
  {https://ui.adsabs.harvard.edu/abs/2011Natur.474..616M} {474, 616}

\bibitem[\protect\citeauthoryear{Myers et~al.,}{Myers et~al.}{2015}]{Myers2015}
Myers A.~D.,  et~al., 2015, \mn@doi [The Astrophysical Journal Supplement
  Series] {10.1088/0067-0049/221/2/27}, 221, 27

\bibitem[\protect\citeauthoryear{{Nanni}, {Hennawi}, {Wang}, {Yang},
  {Schindler}  \& {Fan}}{{Nanni} et~al.}{2022}]{Nanni2022}
{Nanni} R.,  {Hennawi} J.~F.,  {Wang} F.,  {Yang} J.,  {Schindler} J.-T.,
  {Fan} X.,  2022, \mn@doi [\mnras] {10.1093/mnras/stac1944}, \href
  {https://ui.adsabs.harvard.edu/abs/2022MNRAS.515.3224N} {515, 3224}

\bibitem[\protect\citeauthoryear{Paszke et~al.,}{Paszke et~al.}{2019}]{pytorch}
Paszke A.,  et~al., 2019, in , Advances in Neural Information Processing
  Systems 32.
Curran Associates, Inc., pp 8024--8035, \url
  {http://papers.neurips.cc/paper/9015-pytorch-an-imperative-style-high-performance-deep-learning-library.pdf}

\bibitem[\protect\citeauthoryear{{Ritchie} \& {Murray}}{{Ritchie} \&
  {Murray}}{2019}]{sXD}
{Ritchie} J.~A.,  {Murray} I.,  2019, arXiv e-prints, \href
  {https://ui.adsabs.harvard.edu/abs/2019arXiv191111663R} {p. arXiv:1911.11663}

\bibitem[\protect\citeauthoryear{Robbins \& Monro}{Robbins \&
  Monro}{1951}]{SDG1}
Robbins H.,  Monro S.,  1951, \mn@doi [The Annals of Mathematical Statistics]
  {10.1214/aoms/1177729586}, 22, 400

\bibitem[\protect\citeauthoryear{{Schlafly}, {Meisner}  \& {Green}}{{Schlafly}
  et~al.}{2019}]{Schlafly2019}
{Schlafly} E.~F.,  {Meisner} A.~M.,   {Green} G.~M.,  2019, \mn@doi [\apjs]
  {10.3847/1538-4365/aafbea}, \href
  {https://ui.adsabs.harvard.edu/abs/2019ApJS..240...30S} {240, 30}

\bibitem[\protect\citeauthoryear{Stefanski \& Carroll}{Stefanski \&
  Carroll}{1990}]{Stefanski1990}
Stefanski L.~A.,  Carroll R.~J.,  1990, \mn@doi [Statistics]
  {10.1080/02331889008802238}, 21, 169

\bibitem[\protect\citeauthoryear{Tabak \& Turner}{Tabak \&
  Turner}{2013}]{Tabak2013}
Tabak E.,  Turner C.,  2013, \mn@doi [Communications on Pure and Applied
  Mathematics] {10.1002/cpa.21423}, 66, 145

\bibitem[\protect\citeauthoryear{Tabak \& Vanden-Eijnden}{Tabak \&
  Vanden-Eijnden}{2010}]{Tabak2010}
Tabak E.,  Vanden-Eijnden E.,  2010, \mn@doi [Communications in Mathematical
  Sciences] {10.4310/CMS.2010.v8.n1.a11}, 8, 217

\bibitem[\protect\citeauthoryear{{Wang} et~al.,}{{Wang}
  et~al.}{2018}]{Wang2018}
{Wang} F.,  et~al., 2018, \mn@doi [\apjl] {10.3847/2041-8213/aaf1d2}, \href
  {https://ui.adsabs.harvard.edu/abs/2018ApJ...869L...9W} {869, L9}

\bibitem[\protect\citeauthoryear{{Wang} et~al.,}{{Wang}
  et~al.}{2019}]{Wang2019}
{Wang} F.,  et~al., 2019, \mn@doi [\apj] {10.3847/1538-4357/ab2be5}, \href
  {https://ui.adsabs.harvard.edu/abs/2019ApJ...884...30W} {884, 30}

\bibitem[\protect\citeauthoryear{{Wang} et~al.,}{{Wang}
  et~al.}{2021}]{Wang_2021}
{Wang} F.,  et~al., 2021, \mn@doi [\apjl] {10.3847/2041-8213/abd8c6}, \href
  {https://ui.adsabs.harvard.edu/abs/2021ApJ...907L...1W} {907, L1}

\bibitem[\protect\citeauthoryear{White et~al.,}{White et~al.}{2012}]{White2012}
White M.,  et~al., 2012, \mn@doi [Monthly Notices of the Royal Astronomical
  Society] {10.1111/j.1365-2966.2012.21251.x}, 424, 933

\bibitem[\protect\citeauthoryear{{Wolfson}, {Hennawi}, {Davies}  \&
  {O{\~n}orbe}}{{Wolfson} et~al.}{2023}]{Wolfson2023}
{Wolfson} M.,  {Hennawi} J.~F.,  {Davies} F.~B.,   {O{\~n}orbe} J.,  2023,
  \mn@doi [\mnras] {10.1093/mnras/stad701}, \href
  {https://ui.adsabs.harvard.edu/abs/2023MNRAS.tmp..677W} {}

\bibitem[\protect\citeauthoryear{{Yang} et~al.,}{{Yang}
  et~al.}{2019}]{Yang2019}
{Yang} J.,  et~al., 2019, \mn@doi [\aj] {10.3847/1538-3881/ab1be1}, \href
  {https://ui.adsabs.harvard.edu/abs/2019AJ....157..236Y} {157, 236}

\bibitem[\protect\citeauthoryear{{Yang} et~al.,}{{Yang}
  et~al.}{2020}]{Yang_2020}
{Yang} J.,  et~al., 2020, \mn@doi [\apj] {10.3847/1538-4357/abbc1b}, \href
  {https://ui.adsabs.harvard.edu/abs/2020ApJ...904...26Y} {904, 26}

\bibitem[\protect\citeauthoryear{{York} et~al.,}{{York}
  et~al.}{2000}]{York_SDSS}
{York} D.~G.,  et~al., 2000, \mn@doi [\aj] {10.1086/301513}, \href
  {https://ui.adsabs.harvard.edu/abs/2000AJ....120.1579Y} {120, 1579}

\bibitem[\protect\citeauthoryear{Zhang}{Zhang}{1990}]{Zhang1990}
Zhang C.-H.,  1990, \mn@doi [The Annals of Statistics]
  {10.1214/aos/1176347627}, 18, 806

\makeatother
\end{thebibliography}




\appendix

\section{Density Distribution and Contours}

\begin{figure*}
    \includegraphics[width=\textwidth]{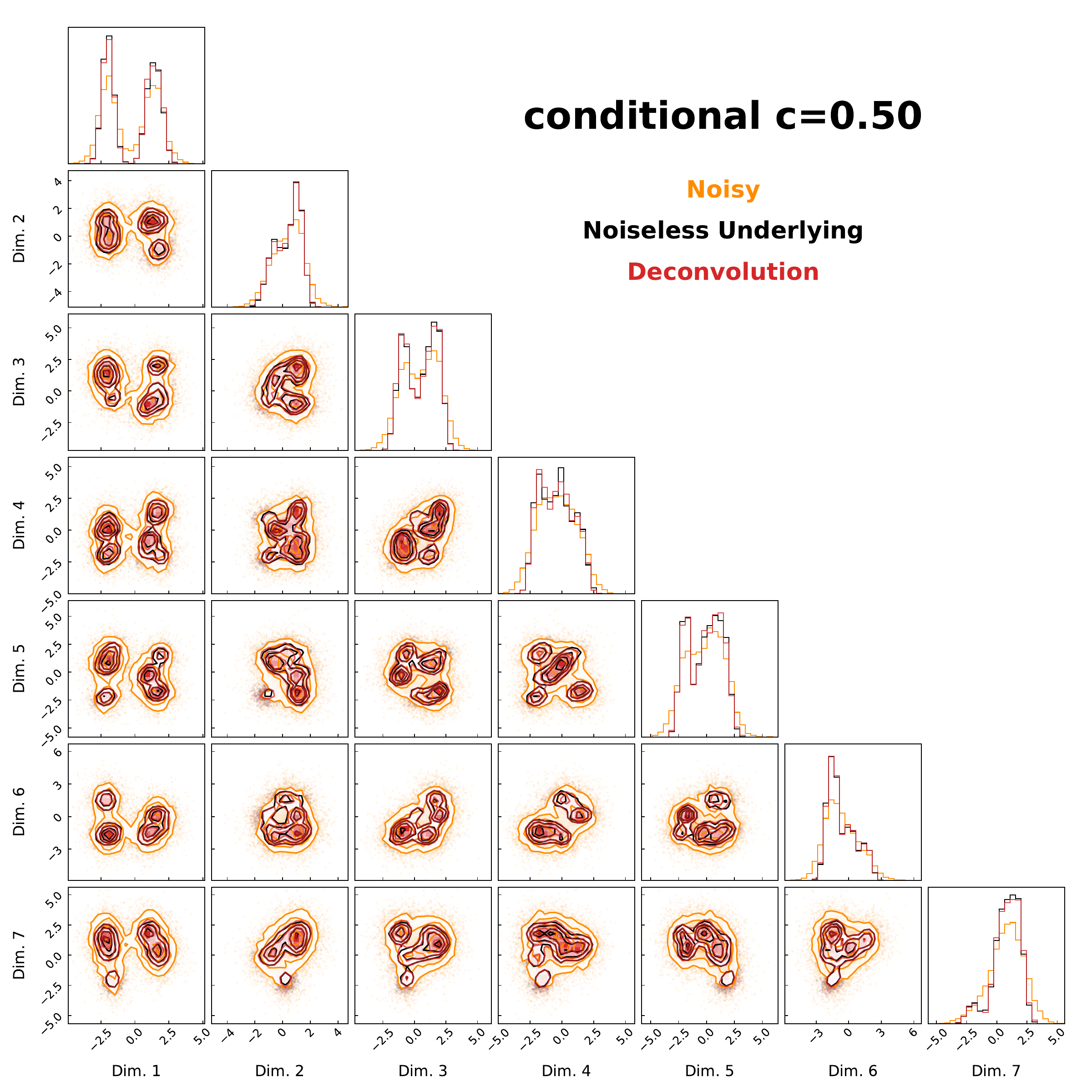}
    \caption{The distribution and density contours of $10,000$ samples from each of the noisy toy model, the underlying toy model and the deconvolution when $\mathbf{c}=0.90$. On the upper or right panels show the 1-D marginal distribution of the samples. Color scheme is the same as Fig \ref{fig:Comp_0.1} and \ref{fig:Comp_0.9}.}
    \label{fig:Comp_0.5}
\end{figure*}


\bsp	
\label{lastpage}
\end{document}